\documentclass[10pt,twocolumn,aps,prd,preprintnumbers,showpacs,superscriptaddress,nofootinbib,amsmath,amssymb,floats,floatfix,showkeys,notitlepage,longbibliography]{revtex4-2} 
\usepackage{orcidlink}
\usepackage{comment}
\usepackage{lipsum}
\usepackage{graphicx}
\usepackage{subfigure}
\usepackage{palatino}
\usepackage{sans}
\usepackage{hyperref}
\hypersetup{colorlinks=true,linkcolor=blue,urlcolor=blue,citecolor=blue}
\usepackage[toc,page]{appendix}
\usepackage[normalem]{ulem}
\usepackage{adjustbox}
\usepackage{latexsym}
\usepackage{amsmath}
\usepackage{amssymb}
\usepackage{amsfonts}
\usepackage{dcolumn}
\usepackage{bm}
\usepackage{tikz}
\usetikzlibrary{decorations.pathmorphing}
\usepackage{bigints}
\usepackage{array,tabularx,multirow,booktabs}
\usepackage[tracking=true]{microtype}
\usepackage{soul} 
\SetTracking{}{500}
\SetTracking{encoding={*}, shape=sc}{40}
\UseRawInputEncoding 
\allowdisplaybreaks
\usepackage[utf8]{inputenc}
\usepackage{xcolor} 

\begin{document} \sloppy

\title{Reference-renormalized curvature-primitive Gauss-Bonnet formalism for finite-distance weak gravitational lensing in static spherical spacetimes}

\author{Reggie C. Pantig \orcidlink{0000-0002-3101-8591}} 
\email{rcpantig@mapua.edu.ph}
\affiliation{Physics Department, School of Foundational Studies and Education, Map\'ua University, 658 Muralla St., Intramuros, Manila 1002, Philippines.}

\author{Ali \"Ovg\"un \orcidlink{0000-0002-9889-342X}}
\email{ali.ovgun@emu.edu.tr}
\affiliation{Physics Department, Eastern Mediterranean University, Famagusta, 99628 North
Cyprus via Mersin 10, Turkiye.}

\begin{abstract}
We develop a reference-renormalized (photon-sphere-free) normalization scheme for Gauss-Bonnet gravitational lensing at finite distance in static, spherically symmetric spacetimes. The method treats the curvature primitive used to reduce the Gauss-Bonnet curvature-area integral as a quantity defined only modulo an additive constant (an additive gauge freedom). We fix this gauge by matching to a physically chosen reference optical geometry in an outer regime where the physical geometry approaches that reference, thereby defining a unique renormalized discrepancy primitive $\mathcal{P}_e(r)$ by reference subtraction. The resulting master formula yields the Ishihara-Li finite-distance deflection angle without invoking any circular null orbit, while remaining fully compatible with orbit-normalized prescriptions whenever a suitable photon sphere exists (the two gauges differ only by a constant shift and give identical $\alpha$). In asymptotically flat settings the canonical reference is Minkowski, while in Kottler-type backgrounds the canonical reference is de Sitter within the static patch, making the operational fiducial explicit. We validate the method by reproducing Ishihara's finite-distance weak-deflection formulas for Schwarzschild, Reissner-Nordstr\"om, and Kottler spacetimes, including the mixed $r_g\Lambda$ term in the Kottler case within the static-patch fiducial. We also present a demonstrative example in which orbit normalization is genuinely inapplicable because no circular null orbit exists in the physical optical region (the Janis-Newman-Winicour spacetime for $\gamma\le \tfrac12$). The result is a unified, geometrically transparent route to finite-distance lensing that preserves compatibility with orbit-normalized prescriptions whenever those apply.
\end{abstract}


\maketitle

\section{Introduction}\label{sec1}
Gravitational deflection of light remains one of the most informative probes of spacetime geometry, spanning regimes from weak-field solar-system tests to strong-field imaging and lensing by compact objects \cite{Schneider:1992bmb,Perlick:2004tq}. While the textbook derivation of the bending angle in asymptotically flat spacetimes is well understood, modern applications increasingly demand definitions and schemes that (i) remain meaningful when the source and receiver are at finite distances and (ii) retain geometric meaning in backgrounds that are not asymptotically Euclidean \cite{Ishihara:2016vdc,Arakida:2017hrm,Takizawa:2020egm}. In non-asymptotically flat spacetimes, the notion of a straight reference trajectory is intrinsically background-dependent, and naive infinite-distance limits may be ill-defined or physically irrelevant \cite{Takizawa:2020egm,Arakida:2017hrm,Gibbons:2008ru,Rindler:2007zz}.

A particularly influential geometric framework is the Gauss-Bonnet approach to lensing in optical geometry, initiated in its modern form by Gibbons and Werner \cite{Gibbons:2008rj}. In this viewpoint, the bending angle is encoded in global geometric data of a domain in the optical manifold, with curvature entering through a surface integral constrained by topology. Its earliest implementations, however, were tailored to asymptotic lensing in effectively Euclidean regions, where both the definition of the deflection angle and the treatment of boundary terms simplify.

A decisive step toward a finite-distance, operationally invariant definition in static, spherically symmetric geometries was provided by Ishihara and collaborators \cite{Ishihara:2016vdc,Ishihara:2016sfv}. They defined a finite-distance deflection angle in terms of locally measured endpoint angles and a coordinate separation angle, and justified its invariance using the Gauss-Bonnet theorem on the optical manifold \cite{Ishihara:2016vdc}. This clarified that the Gauss-Bonnet method is not merely a calculational shortcut: it supplies a precise definition of the observable that remains meaningful when endpoints are not at infinity \cite{Ishihara:2016vdc,Ono:2019hkw}. Subsequent work systematized finite-distance corrections and their interpretation \cite{Ono:2019hkw,Takizawa:2020egm,Arakida:2017hrm,Ishihara:2016sfv}.

For explicit evaluations, one often seeks to reduce the curvature-area term to a boundary functional by expressing the curvature density as a total radial derivative. This curvature-primitive strategy replaces a two-dimensional integral by a one-dimensional functional along the ray together with endpoint contributions, and it can greatly simplify weak-deflection calculations in static spherical geometries. The primitive, however, is not unique: like any antiderivative it is defined only up to an additive constant. In asymptotically flat geometries, a canonical representative is selected by matching to the Euclidean (Minkowski) reference at large radius. In more general settings, the issue is not physical dependence on the additive constantbut the absence of an intrinsic, closure-independent way to fix a representative before taking limits or making weak-field replacements. The exact curvature contribution depends only on primitive differences, and this is the normalization gap addressed here.

Li and collaborators introduced a curvature-primitive reformulation of Gauss-Bonnet lensing in which the Gaussian-curvature term is reduced to a boundary functional by introducing a primitive $\mathcal{P}(r)$ of the curvature density. The structural point is that such primitives carry an \emph{additive gauge freedom}. The exact Gauss-Bonnet/primitive identity is gauge invariant because it depends only on boundary differences, but practical weak-deflection work (e.g. evaluating along a reference ray) still requires a consistent gauge fixing to prevent spurious endpoint-dependent constants from contaminating intermediate expressions. Orbit normalization at a circular null orbit (photon sphere) is one convenient choice in standard black-hole spacetimes, but it is not universal: circular null orbits may be absent in the physical optical region, may lie outside the static manifold region on which the Gauss-Bonnet domain is constructed, or may be misaligned with finite-distance observables defined by endpoint angle comparisons \cite{Li:2020wvn,Claudel:2000yi,Sau:2020xau}. This approach \cite{Gibbons:2008rj,Ishihara:2016vdc,Li:2020wvn} has inspired a large body of work across alternative theories of gravity, and effective media settings such as magnetic fields, plasma, and even dark matter \cite{Huang:2025mek,Ghosh:2025jpb,Qiao:2025ojr,Huang:2025vqm,Lu:2025mcm,Mushtaq:2025ksk,Wang:2025cyd,Pantig:2024asu,Pantig:2024kqy,Gao:2024ejs,Liu:2024wal,Zhang:2024uex,Ovgun:2025ctx,Soares:2025hpy,Rincon:2024won,Li:2024ujw,Liu:2023xtb,Gao:2023ltr}.

We therefore fix the additive gauge by matching to a physically chosen reference optical geometry in an outer regime where the physical geometry approaches that reference. Concretely, we define a renormalized primitive by subtracting the primitive associated with an admissible reference optical geometry and by fixing the additive ambiguity through a matching condition in the overlap region. For asymptotically flat spacetimes, the reference is Minkowski. For Kottler-type backgrounds, the natural reference is de Sitter in the static patch. This choice is not an extra assumption; it is the minimal input required to interpret finite-distance angle comparisons as deflection rather than as coordinate-dependent artifacts. The resulting formalism is photon-sphere-free: it does not require a circular null orbit to exist or to be used as an anchor.

We illustrate the construction in the weak-deflection, finite-distance regime for Schwarzschild, Reissner-Nordstr\"om, and Kottler, and we add a demonstrative case where photon-sphere anchoring fails because no circular null orbit exists outside the curvature singularity (the Janis-Newman-Winicour spacetime in the no-photon-sphere regime $\gamma\le \frac{1}{2}$). Ishihara's finite-distance expressions provide stringent checks for Schwarzschild and Kottler, including the Kottler mixed term that has been central in the broader literature. Our emphasis is methodological and geometric: we restrict to static, spherically symmetric spacetimes and to finite-distance configurations without multiple winding, where the optical manifold description is clean and the Gauss-Bonnet domain can be chosen unambiguously.

After fixing conventions and the finite-distance optical-geometry setup, we develop the reference-renormalized curvature-primitive formulation and apply it to these representative finite-distance weak-deflection examples.

The remainder of this paper is structured as follows. Section \ref{sec2} fixes conventions and reviews the finite-distance deflection
observable and the optical geometry. Section \ref{sec3} presents the finite-distance Gauss-Bonnet relation and its boundary bookkeeping.
Section \ref{sec4} reviews the curvature-primitive reduction and interprets orbit normalization as an additive gauge fixing when a suitable
circular null orbit exists. Section \ref{sec5} introduces the reference-renormalized primitive $\mathcal{P}_e$ and derives the photon-sphere-free master
formula. Section \ref{sec6} summarizes general properties and the domain of validity of the construction. Section \ref{sec7} develops the weak-deflection
expansion strategy in a form suited to finite-distance comparisons. Section \ref{sec8} presents three examples (Schwarzschild,
Reissner-Nordstr\"om, and Kottler) with explicit agreement with Ishihara's finite-distance results, and then adds a demonstrative
no-photon-sphere case (Janis-Newman-Winicour with $\gamma\le\tfrac12$) where orbit normalization is genuinely inapplicable.
Section \ref{sec9} discusses interpretation, including the reference dependence of the Kottler mixed term, and Section \ref{sec10} concludes.

\section{Conventions and geometric setup} \label{sec2}
This section fixes conventions and records the geometric constructions used in what follows.

\subsection{Spacetime class and equatorial reduction} \label{sec2.1}
We work on a four-dimensional, static, spherically symmetric spacetime with coordinates $(t,r,\theta,\phi)$ adapted to the hypersurface-orthogonal Killing field $\partial_t$ and the rotational symmetries. The signature is $(-,+,+,+)$ and we use units $G=c=1$. Greek indices $\mu,\nu,\ldots$ run over $0,1,2,3$ with $x^0=t$, and repeated indices are summed. We restrict to the static region where $A(r)>0$. The metric is written as
\begin{equation}
ds^2=-A(r)\,dt^2+B(r)\,dr^2+C(r)d\Omega^2,
\label{2.1}
\end{equation}
with $A(r),B(r),C(r)>0$ in the region of interest, and $d\Omega^2 = d\theta^2+\sin^2\theta\,d\phi^2$; $C(r)=r^2$ corresponds to the areal radius, but we keep $C(r)$ arbitrary within the spherical class.

Null rays are affinely parametrized null geodesics $x^\mu(\lambda)$ with $ds^2=0$. Stationarity and axisymmetry give the conserved quantities
\begin{equation}
E\equiv -p_t=A(r)\,\frac{dt}{d\lambda},\qquad L\equiv p_\phi=C(r)\sin^2\theta\,\frac{d\phi}{d\lambda},
\label{2.2}
\end{equation}
where $p_\mu=g_{\mu\nu}\,dx^\nu/d\lambda$. By spherical symmetry we may choose the orbital plane to be equatorial: $\theta=\pi/2$ and $d\theta/d\lambda=0$, so $\sin^2\theta=1$ and $L=C(r)\,d\phi/d\lambda$. We also define the impact parameter
\begin{equation}
b \equiv \frac{|L|}{E} \ge 0 .
\label{2.3}
\end{equation}
For future-directed rays we take $E>0$; the sign of $L$ fixes the azimuthal orientation and will not enter explicitly.

Applying the null condition to Eq. \eqref{2.1} with $\theta=\pi/2$ yields the radial equation
\begin{equation}
B(r)\left(\frac{dr}{d\lambda}\right)^2=\frac{E^2}{A(r)}-\frac{L^2}{C(r)}.
\label{2.4}
\end{equation}
A turning point $r=r_0$ (when present) satisfies $dr/d\lambda=0$, hence $E^2/A(r_0)=L^2/C(r_0)$ or $b^2=C(r_0)/A(r_0)$. No assumption of a circular null orbit is made at this stage.

\subsection{Optical geometry for null rays} \label{sec2.2}
In a static spacetime, null geodesics project to geodesics of an associated Riemannian optical geometry. Setting $ds^2=0$ in Eq. \eqref{2.1} gives
\begin{equation}
dt^2=\frac{1}{A(r)}\left[B(r)\,dr^2+C(r)\,\left(d\theta^2+\sin^2\theta\,d\phi^2\right)\right].
\label{2.5}
\end{equation}
Along null curves we define $d\sigma^2\equiv dt^2$, so the optical metric on the spatial coordinates is $\bar{g}_{ij}=g_{ij}/A(r)$. Restricting to $\theta=\pi/2$ gives the two-dimensional optical manifold with coordinates $(x^a)=(r,\phi)$ and metric
\begin{equation}
d\sigma^2=\bar{g}_{ab}\,dx^a dx^b=\frac{B(r)}{A(r)}\,dr^2+\frac{C(r)}{A(r)}\,d\phi^2,
\label{2.6}
\end{equation}
and $x^a=(r,\phi)$. For future-directed rays we take $dt>0$, hence $d\sigma=dt$ along the ray. We write $\bar{\nabla}$ for the Levi-Civita connection of $\bar{g}_{ab}$ and denote by $K$ the Gaussian curvature of the optical manifold. The corresponding area element is
\begin{equation}
dS=\sqrt{\det\bar{g}}\,dr\,d\phi=\frac{\sqrt{B(r)\,C(r)}}{A(r)}\,dr\,d\phi.
\label{2.7}
\end{equation}
In Gauss-Bonnet lensing, the curvature contribution enters through integrals of $K\,dS$ over suitable domains in this optical manifold. The construction relies only on the static metric and does not presuppose asymptotic flatness.

\subsection{Finite-distance lensing geometry: endpoints, angles, and observable} \label{sec2.3}
Let a source event $S$ and a receiver (observer) event $R$ be connected by a null geodesic in the equatorial plane. Their spatial positions in the optical manifold are $(r_S,\phi_S)$ and $(r_R,\phi_R)$, and the ray projects to a curve $\gamma$ in $(r,\phi)$ that is an unparameterized geodesic of Eq. \eqref{2.6}.

To define local direction angles, choose an orthonormal frame for $\bar{g}_{ab}$,
\begin{equation}
e_{(r)}=\sqrt{\frac{A(r)}{B(r)}}\,\partial_r,\qquad
e_{(\phi)}=\sqrt{\frac{A(r)}{C(r)}}\,\partial_\phi.
\label{2.8}
\end{equation}
Let $T^a$ be the optical unit tangent to $\gamma$, $\bar{g}_{ab}T^aT^b=1$, using the optical arclength parameter $\sigma$ so that $T^a\equiv dx^a/d\sigma$. The angle $\Psi\in(0,\pi)$ between the ray direction and the outward radial direction is defined by
\begin{equation}
\cos\Psi \equiv \bar{g}_{ab}\,T^a e_{(r)}^{\,b},\qquad
\sin\Psi \equiv \bar{g}_{ab}\,T^a e_{(\phi)}^{\,b},
\label{2.9}
\end{equation}
with the orientation fixed by the sign of increasing $\phi$ along the ray. Using Eq. \eqref{2.2} and $d\sigma=dt$ along null curves, one finds
\begin{equation}
\frac{d\phi}{d\sigma}
=\frac{(d\phi/d\lambda)}{(dt/d\lambda)}
=\frac{(L/C)}{(E/A)}
=\mathrm{sgn}(L)\,b\,\frac{A(r)}{C(r)}.
\end{equation}
Since $e_{(\phi)}$ is unit with respect to $\bar{g}_{ab}$, Eq. \eqref{2.9} gives $\sin\Psi=\sqrt{\bar{g}_{\phi\phi}}\,d\phi/d\sigma$, and with $\bar{g}_{\phi\phi}=C/A$ this yields
\begin{equation}
\sin\Psi(r)= b\,\sqrt{\frac{A(r)}{C(r)}}.
\label{2.10}
\end{equation}
Thus $\Psi_S\equiv\Psi(r_S)$ and $\Psi_R\equiv\Psi(r_R)$ are determined by $(b,r_S)$ and $(b,r_R)$ alone.

The finite-distance deflection angle compares the endpoint directions with the azimuthal separation accumulated along the ray. Let $\phi_{RS}$ be the net change in $\phi$ from $S$ to $R$, taken positive when $\phi$ increases along the propagation direction. We define
\begin{equation}
\alpha \equiv \Psi_R-\Psi_S+\phi_{RS}.
\label{2.11}
\end{equation}
Here $\Psi_S$ and $\Psi_R$ are locally defined angles on the optical manifold, while $\phi_{RS}$ will later be related to curvature on an associated domain. For suitable large-distance limits, Eq. \eqref{2.11} reduces to the standard asymptotic deflection angle.

In the weak-deflection, single-pass configurations considered here (no multiple winding), the ray has a single turning point between $S$ and $R$, so the direction at $S$ is ingoing and at $R$ outgoing relative to the outward radial direction. Accordingly, when Eq. \eqref{2.10} is inverted we adopt the principal branch $\arcsin\in[0,\pi/2]$ and take
\begin{align}
\Psi_R&=\arcsin\!\left(b\sqrt{\frac{A(r_R)}{C(r_R)}}\right),\nonumber \\
\Psi_S&=\pi-\arcsin\!\left(b\sqrt{\frac{A(r_S)}{C(r_S)}}\right),
\label{eq:Psi_branch_weak}
\end{align}
which is the convention used consistently in the explicit evaluations. In strong-deflection or multi-image situations, the branch structure and the associated domain construction must be treated separately.

\section{The finite-distance Gauss-Bonnet relation - a brief review} \label{sec3}
We recall the finite-distance Gauss-Bonnet relation on the equatorial optical manifold, which rewrites the operational deflection angle in Eq. \eqref{2.11} in terms of intrinsic curvature on a suitable domain.

\subsection{Domain construction and Gauss-Bonnet identity} \label{sec3.1}
Let $(\mathcal{M},\bar{g}_{ab})$ be the two-dimensional optical manifold introduced in Eq. \eqref{2.6}, with area element $dS$ from Eq. \eqref{2.7} and Gaussian curvature $K$. For a compact, oriented, simply connected domain $D\subset\mathcal{M}$ with piecewise smooth boundary $\partial D$ and finitely many vertices, the Gauss-Bonnet theorem reads
\begin{equation}
\iint_D K\,dS+\oint_{\partial D}\kappa_g\,d\ell+\sum_{i=1}^{N}\beta_i
=2\pi\,\chi(D),
\label{3.1}
\end{equation}
where $\kappa_g$ is the geodesic curvature along smooth boundary segments, $d\ell$ is the induced optical line element, $\beta_i$ are exterior (jump) angles at the $N$ vertices, and $\chi(D)$ is the Euler characteristic. For the domains used here, $\chi(D)=1$.

To connect Eq. \eqref{3.1} with finite-distance lensing, take $D$ to be the standard quadrilateral domain bounded by: (i) the projected ray $\gamma$ from the source $S=(r_S,\phi_S)$ to the receiver $R=(r_R,\phi_R)$, (ii) the two radial curves $\gamma_S$ and $\gamma_R$ given by $\phi=\phi_S$ and $\phi=\phi_R$, and (iii) a closing curve $C_\Gamma$ at $r=r_\Gamma$ joining the outer endpoints of $\gamma_S$ and $\gamma_R$. With the usual positive orientation, $\gamma$ is a geodesic of $(\mathcal{M},\bar{g}_{ab})$, and $\gamma_S$ and $\gamma_R$ are geodesics as $\phi=\text{const}$ lines in an axisymmetric metric depending only on $r$. Hence $\kappa_g=0$ on $\gamma$, $\gamma_S$, and $\gamma_R$, and the boundary integral in Eq. \eqref{3.1} reduces to the contribution from $C_\Gamma$. Two common choices are: an outer-arc closure in a reference/Euclidean regime, for which $\kappa_g\,d\ell\to d\phi$ and therefore
\begin{equation}
\int_{C_\Gamma}\kappa_g\,d\ell\;\to\;\int_{\phi_R}^{\phi_S} d\phi \;=\;\phi_S-\phi_R,
\label{3.2}
\end{equation}
so that (with the sign convention of Eq. \eqref{2.11}) $\int_{C_\Gamma}\kappa_g\,d\ell\to -\phi_{RS}$; or a geodesic closure, for which $\kappa_g=0$ on $C_\Gamma$.

The two outer vertices (where a radial geodesic meets $C_\Gamma$) contribute a net exterior angle $\pi$ in the reference regime. At the physical endpoints, the exterior angles are fixed by the local propagation angles of Eq. \eqref{2.9}:
\begin{equation}
\beta_S=\Psi_S,
\label{3.3}
\end{equation}
\begin{equation}
\beta_R=\pi-\Psi_R.
\label{3.4}
\end{equation}
Using $\chi(D)=1$ and the vanishing of $\kappa_g$ on $\gamma$, $\gamma_S$, and $\gamma_R$, Eq. \eqref{3.1} reduces to
\begin{equation}
\iint_{D}K\,dS+\int_{C_\Gamma}\kappa_g\,d\ell+\beta_S+\beta_R=\pi.
\label{3.5}
\end{equation}
With $\beta_S=\Psi_S$ and $\beta_R=\pi-\Psi_R$, Eq. \eqref{3.5} gives
\begin{equation}
\Psi_R-\Psi_S=- \iint_{D} K\, dS \;+\; \int_{C_\Gamma}\kappa_g\, d\ell .
\label{3.6}
\end{equation}

Substituting Eq. \eqref{3.6} into the operational definition \eqref{2.11} yields
\begin{equation}
\alpha=- \iint_{D} K\, dS \;+\; \int_{C_\Gamma}\kappa_g\, d\ell \;+\; \phi_{RS}.
\label{3.7}
\end{equation}
Equation \eqref{3.7} holds for any admissible choice of the outer closure curve $C_\Gamma$. For an outer-arc closure in a reference/Euclidean regime, $\kappa_g\,d\ell \to d\phi$ on $C_\Gamma$, so $\int_{C_\Gamma}\kappa_g\,d\ell = \phi_S-\phi_R = -\phi_{RS}$ and $\alpha = -\iint_D K\,dS$; in particular, for exact Minkowski optical geometry $K=0$ and $\alpha=0$. If $C_\Gamma$ is a geodesic, then $\kappa_g=0$ and Eq. \eqref{3.7} reduces accordingly.

For axisymmetric two-metrics of the form \eqref{2.6}, writing $\bar{g}_{ab}=\mathrm{diag}(\bar{g}_{rr}(r),\bar{g}_{\phi\phi}(r))$, the Gaussian curvature admits the radial-derivative form
\begin{equation}
K
=-\frac{1}{\sqrt{\det\bar{g}}}\frac{d}{dr}\!\left(\frac{\sqrt{\det\bar{g}}}{\bar{g}_{rr}}\,\bar{\Gamma}^{\phi}{}_{r\phi}\right),
\quad
\bar{\Gamma}^{\phi}{}_{r\phi}=\frac{1}{2}\frac{d}{dr}\ln\bar{g}_{\phi\phi}.
\label{3.8}
\end{equation}
Together with $dS=\sqrt{\det\bar{g}}\,dr\,d\phi$ from Eq. \eqref{2.7}, this highlights that $K\,dS$ is controlled by radial structure alone. In the weak-deflection regime one may evaluate Eq. \eqref{3.7} by approximating the ray in the chosen reference geometry while keeping curvature corrections in $K$, with endpoint angles fixed geometrically by Eq. \eqref{2.10}.

\section{Curvature primitive and the normalization issue - Li \emph{et al}. circular orbit method for GBT} \label{sec4}
We review the device that makes Eq. \eqref{3.7} efficient in static spherical settings: rewriting the curvature-area term as an integral of a radial curvature primitive. We then isolate the point relevant to this work: the primitive is defined only up to an additive constant, and the customary way of fixing it in the literature often uses a circular null orbit (photon sphere). We summarize only what is needed; detailed derivations are in Ref. \cite{Li:2020wvn}.

\subsection{Curvature-primitive reduction in the Li formalism} \label{sec4.1}
For the optical metric \eqref{2.6}, the curvature density
\begin{equation}
\mathcal{D}(r)\equiv K(r)\,\sqrt{\det\bar{g}(r)},
\label{4.1}
\end{equation}
depends only on $r$. In the weak-deflection regime, parameterize the ray by $\phi$ so that $r=r(\phi)$ from $\phi_S$ to $\phi_R$. In the finite-distance quadrilateral domain of Section \ref{sec3}, each radial line at fixed $\phi$ intersects the domain between $r=r(\phi)$ and the outer closure radius $r=r_\Gamma$. Hence the curvature term in Eq. \eqref{3.7} decomposes as
\begin{align}
\iint_D K\,dS&=
\int_{\phi_S}^{\phi_R}\left(\int_{r(\phi)}^{r_\Gamma} K(r)\,\sqrt{\det\bar{g}(r)}\,dr\right)d\phi \nonumber \\
&= \int_{\phi_S}^{\phi_R}\left(\int_{r(\phi)}^{r_\Gamma} \mathcal{D}(r)\,dr\right)d\phi.
\label{4.2}
\end{align}
Introduce a radial primitive $\mathcal{P}(r)$ satisfying
\begin{equation}
\frac{d\mathcal{P}}{dr}=\mathcal{D}(r).
\label{4.3}
\end{equation}
Then the inner integral becomes
\begin{equation}
\int_{r(\phi)}^{r_\Gamma}\mathcal{D}(r)\,dr=\mathcal{P}(r_\Gamma)-\mathcal{P}(r(\phi)),
\label{4.4}
\end{equation}
and the curvature-area term reduces to
\begin{equation}
\iint_D K\,dS=
\int_{\phi_S}^{\phi_R}\left[\mathcal{P}(r_\Gamma)-\mathcal{P}(r(\phi))\right]\,d\phi.
\label{4.5}
\end{equation}
The only ambiguity is additive: Eq. \eqref{4.3} fixes $\mathcal{P}(r)$ up to $\mathcal{P}(r)\mapsto \mathcal{P}(r)+{\rm const}$. Because Eq. \eqref{4.5} involves only differences, $\iint_DK\,dS$ is invariant; we interpret the constant shift as an \emph{additive gauge freedom}. Choosing a representative (e.g. imposing $\mathcal{P}(r_\Gamma)=0$ on a chosen closure curve, or fixing $\mathcal{P}$ at a particular radius) is therefore a \emph{gauge choice}. In the circular-orbit prescription of Li \emph{et al}., one sets $r_\Gamma=r_{\rm co}$ and normalizes so that $\mathcal{P}(r_{\rm co})=0$, eliminating the explicit boundary term $\mathcal{P}(r_\Gamma)$ in Eq. \eqref{4.5} and leaving an integrand evaluated along the ray.

\subsection{Orbit normalization as a gauge choice (and its domain of applicability)} \label{sec4.2}
In the implementation emphasized by Li \emph{et al}., the additive freedom in $\mathcal{P}(r)$ is fixed by requiring the primitive to vanish at a distinguished radius. For null lensing in static spherical spacetimes, the distinguished radius is taken to be that of a circular null geodesic, when such an orbit exists and lies within the optical region. Denoting it by $r_\text{co}$,
\begin{equation}
\mathcal{P}(r_\text{co})=0.
\label{4.6}
\end{equation}
This choice is convenient: in common asymptotically flat black-hole metrics, $r_\text{co}$ is intrinsic and typically unique, and it converts $\mathcal{P}(r_\Gamma)$ and $\mathcal{P}(r(\phi))$ into definite integrals anchored at $r_\text{co}$, yielding compact weak-deflection expressions.

The limitation is equally clear: Eq. \eqref{4.6} is available only when an appropriate circular null orbit exists and is usable as a normalization anchor for the Gauss-Bonnet domain. Three obstructions are distinct. (i) Some static spherical geometries admit no circular null orbit in the Riemannian optical region where the finite-distance domain $D$ is defined, so Eq. \eqref{4.6} cannot be imposed. (ii) Even if circular null orbits exist, they may lie behind horizons or outside the static patch supporting the optical manifold used in the construction; then the orbit is not part of the region over which $K\,dS$ is evaluated, and the normalization imports external geometric input not tied to the observable \eqref{2.11}. (iii) In non-asymptotically flat backgrounds, fixing the additive constant by a local orbit does not by itself specify how bending should be referenced relative to an appropriate background optical geometry, whereas the finite-distance observable $\alpha$ is defined by endpoint direction comparisons plus accumulated angular separation.

Accordingly, while Eq. \eqref{4.5} is the essential reduction, orbit normalization in the form \eqref{4.6} is not a universal gauge fixing for finite-distance lensing. We therefore retain the primitive reduction but replace the orbit-based fixing of the additive constant by a normalization anchored to a suitable reference regime consistent with the finite-distance definition of the deflection angle.

\section{New formalism: the reference-renormalized curvature primitive} \label{sec5}
We replace the orbit-based normalization in Eq. \eqref{4.6} by a normalization fixed through a chosen reference optical geometry. The curvature primitive of Section \ref{sec4} is defined only up to an additive constant; we fix this constant by matching to a designated reference in a regime where the physical geometry approaches that reference. This yields a photon-sphere-free normalization while preserving the finite-distance meaning of Eq. \eqref{2.11} and the Gauss-Bonnet relation \eqref{3.7}.

\subsection{Reference optical geometry and admissible matching conditions} \label{sec5.1}
Let the physical spacetime be given by Eq. \eqref{2.1} with $A(r)>0$ on the static region of interest. A reference spacetime is another static, spherically symmetric metric in the same chart,
\begin{equation}
ds_\text{ref}^{2}=-A_\text{ref}(r)\,dt^{2}+B_\text{ref}(r)\,dr^{2}+C_\text{ref}(r)\,d\Omega^2,
\label{5.1}
\end{equation}
with $A_\text{ref}(r)>0$ and $B_\text{ref}(r)>0$. A closely related reference-geometry viewpoint in the Gauss-Bonnet approach (used to clarify the fiducial notion of
straight propagation and the definition/interpretation of the total deflection angle) was discussed by Arakida
\cite{Arakida:2017hrm}.
Our use of a reference geometry is complementary: here it is employed to fix the additive ambiguity of the curvature
primitive by reference subtraction, yielding a unique renormalized discrepancy primitive $\mathcal{P}_e$ without requiring a
circular null orbit.
For asymptotically flat problems the natural reference is Minkowski, while for Kottler-type situations the appropriate
background is de Sitter in the static patch.

The radial coordinate itself has no invariant meaning across different spacetimes.
What is invariantly identified in the present construction is the \emph{symmetry 2-sphere} on which the subtraction is performed.
In spherical symmetry such spheres are canonically labeled by the areal radius
$\rho\equiv \sqrt{C(r)}$ (and $\rho\equiv \sqrt{C_{\rm ref}(r_{\rm ref})}$ in the reference spacetime).
Thus the matching prescription is to compare physical and reference optical data at equal $\rho$.
In computations we often adopt the areal-radius gauge (so that $C(r)=r^2$ and likewise for the reference),
in which case the invariant identification $\rho_{\rm ref}=\rho$ reduces to the notationally simple $r_{\rm ref}=r$.
No physical assumption is being made: one may equivalently keep distinct radial parameters and introduce an explicit map
$r_{\rm ref}=f(r)$ defined by $\sqrt{C_{\rm ref}(r_{\rm ref})}=\sqrt{C(r)}$, or rewrite the one-dimensional integrals in the
common invariant variable $\rho$.

It is useful to separate two distinct issues that can otherwise be conflated.
(i) The \emph{curvature primitive} in a curvature-primitive reduction is defined only up to an additive constant, and this is a
true (additive) gauge freedom. The purpose of the present reference-renormalized construction is precisely to fix that
additive ambiguity canonically by an \emph{endpoint/matching} condition (e.g.\ $\mathcal{P}_e(\infty)=0$ in asymptotically flat
cases, or $\mathcal{P}_e(r_{\rm ref})=0$ in a finite static patch). Once a reference is specified, this condition selects a unique
representative of the primitive class and removes any need for orbit-based normalization. (ii) The \emph{choice of reference optical geometry} is conceptually different in a sense that in non-asymptotically Euclidean settings there is
no unique global notion of an \textit{unbent} trajectory, so a finite-distance deflection angle necessarily carries an (implicit or
explicit) fiducial prescription. Our use of a reference geometry makes that fiducial explicit. In the examples studied
here we adopt the canonical \textit{no-lens} member of the same family: Minkowski for asymptotically flat spacetimes, and for
Kottler we take the de Sitter member obtained by setting $r_g\to 0$ at fixed $\Lambda$ within the same static patch and the
same static-observer operational setup. This choice ensures that the no-lens limit is implemented transparently in the
finite-distance observable and that the $\Lambda$-sensitive background pieces are anchored consistently (see also
Sec. \ref{sec9.2} and the explicit Kottler discussion in Sec. \ref{sec8.3}).

Finally, we emphasize that introducing a reference amounts to an exact decomposition
$\mathcal{D}=\mathcal{D}_{\rm ref}+\Delta\mathcal{D}$ inside the Gauss-Bonnet identity: changing $\mathcal{D}_{\rm ref}$ reshuffles
terms between the explicit reference contribution and the discrepancy primitive, but does not alter the underlying
finite-distance deflection angle $\alpha$ (any residual dependence can only appear after weak-field truncation).

Restricting Eq. \eqref{5.1} to null curves and to $\theta=\pi/2$ gives the equatorial reference optical metric
\begin{equation}
d\sigma_\text{ref}^{2}
=\frac{B_\text{ref}(r)}{A_\text{ref}(r)}\,dr^{2}+\frac{C_\text{ref}(r)}{A_\text{ref}(r)}\,d\phi^{2},
\label{5.2}
\end{equation}
with area element
\begin{equation}
dS_\text{ref}
=\sqrt{\det\bar{g}_\text{ref}}\,dr\,d\phi
=\frac{\sqrt{B_\text{ref}(r)\,C_\text{ref}(r)}}{A_\text{ref}(r)}\,dr\,d\phi,
\label{5.3}
\end{equation}
Gaussian curvature $K_\text{ref}$, and curvature density
\begin{equation}
\mathcal{D}_\text{ref}(r)\equiv K_\text{ref}(r)\,\sqrt{\det\bar{g}_\text{ref}(r)}.
\label{5.4}
\end{equation}
These mirror Eqs. \eqref{2.6}, \eqref{2.7}, and \eqref{4.1} for the physical optical manifold.

We assume a matching regime $\mathcal{I}_\text{match}$ where both optical metrics are regular and the physical geometry is well approximated by the reference. This regime is an end of the optical manifold: for asymptotically flat spacetimes one may take $\mathcal{I}_\text{match}=(R,\infty)$, while for a static patch one may take $\mathcal{I}_\text{match}=(R,r_\text{h})$ with $R$ well inside the region where $A(r)>0$. In the static-patch case, endpoints and the relevant ray segment must lie where static observers exist and the optical metric is Riemannian.

We introduce a subtraction radius $r_{\rm ref}\in\mathcal{I}_\text{match}$ that labels the outer 2-sphere where the additive gauge will be fixed. To specify this invariantly within the spherical class, we use its areal radius
\begin{equation}
\rho_{\rm ref}\equiv \sqrt{C(r_{\rm ref})},
\label{eq:areal_rref}
\end{equation}
so $\rho_{\rm ref}=r_{\rm ref}$ when $C(r)=r^2$. Changing $\rho_{\rm ref}$ shifts individual primitives by constants; the exact finite-distance deflection is insensitive to this shift, while any residual $\rho_{\rm ref}$-dependence after weak-field truncation is of higher order than the retained terms.

The normalization is anchored by the integrability of the curvature-density discrepancy
\begin{equation}
\Delta\mathcal{D}(r)\equiv \mathcal{D}(r)-\mathcal{D}_\text{ref}(r),
\label{5.5}
\end{equation}
where $\mathcal{D}(r)=K(r)\sqrt{\det\bar{g}(r)}$ is defined in Eq. \eqref{4.1}. We require that the improper integral of $\Delta\mathcal{D}$ converge toward the matching end. Concretely, for each finite $r$ in the region of interest we impose
\begin{equation}
\int_{r}^{\infty}\left|\Delta\mathcal{D}(u)\right|\,du<\infty,
\label{5.6}
\end{equation}
in the asymptotically flat case, and
\begin{equation}
\int_{r}^{r_{\rm ref}}\left|\Delta\mathcal{D}(u)\right|\,du<\infty
\label{5.7}
\end{equation}
when the matching end is finite. These are the minimal conditions needed to define a reference-normalized discrepancy primitive without invoking a distinguished null orbit.

When we also represent the \emph{reference} curvature-area term by a reference-normalized primitive, we assume integrability of $\mathcal{D}_{\rm ref}$ toward the same endpoint:
\begin{align}
\int_r^{\infty}& |\mathcal{D}_{\rm ref}(u)|\,du < \infty
\quad\text{(asymptotically flat reference)},
\nonumber \\
\int_r^{r_{\rm ref}}& |\mathcal{D}_{\rm ref}(u)|\,du < \infty
\quad\text{(finite-endpoint reference)}.
\label{V.ref_adm}
\end{align}
If this fails, the reference contribution is best kept in boundary/operational form, while the discrepancy primitive remains well defined under \eqref{5.6}-\eqref{5.7}. A concrete example is the de Sitter reference used later for the Kottler spacetime.  With
$A_{\rm dS}(r)=1-\Lambda r^2/3$, $B_{\rm dS}(r)=A_{\rm dS}^{-1}$, $C_{\rm dS}(r)=r^2$, the equatorial optical metric is
\begin{equation}
d\ell^2_{\rm dS}=\bar g^{\rm dS}_{rr}\,dr^2+\bar g^{\rm dS}_{\phi\phi}\,d\phi^2
=\frac{dr^2}{A_{\rm dS}(r)^2}+\frac{r^2}{A_{\rm dS}(r)}\,d\phi^2.
\end{equation}
Its Gaussian curvature is constant,
$K_{\rm dS}=-\Lambda/3$, and the corresponding curvature density is
\begin{equation}
\mathcal{D}_{\rm dS}(r)\equiv K_{\rm dS}\sqrt{\det\bar g_{\rm dS}}
=-\frac{\Lambda}{3}\,\frac{r}{A_{\rm dS}(r)^{3/2}}.
\end{equation}
Hence $\int^{r_h}\!|\mathcal{D}_{\rm dS}(u)|\,du$ diverges as $r\to r_h=\sqrt{3/\Lambda}$, so it is natural in the static patch to keep the
\emph{reference} contribution in boundary/operational form (endpoint angles and the orbit span), while applying the
reference-normalized primitive construction to the \emph{discrepancy} $\Delta \mathcal{D}$ in the overlap regime.

Conditions \eqref{5.6}-\eqref{5.7} do not require pointwise agreement of the metric coefficients; they require only that the integrated curvature discrepancy be well defined. Once the reference geometry is fixed for the physical problem, no further orbit-based prescription is needed to fix the additive gauge of the primitive.

\subsection{Definition of the renormalized primitive} \label{sec5.2}
We define the reference-renormalized curvature primitive that replaces orbit normalization. The physical primitive $\mathcal{P}(r)$ is any function satisfying
\begin{equation}
\frac{d\mathcal{P}}{dr}=\mathcal{D}(r),
\label{5.8}
\end{equation}
with $\mathcal{D}(r)$ from Eq. \eqref{4.1}. Similarly, a reference primitive $\mathcal{P}_\text{ref}(r)$ satisfies
\begin{equation}
\frac{d\mathcal{P}_\text{ref}}{dr}=\mathcal{D}_\text{ref}(r),
\label{5.9}
\end{equation}
with $\mathcal{D}_\text{ref}(r)$ from Eq. \eqref{5.4}. Each primitive is defined only up to an additive constant, but their difference is governed by the integrable discrepancy \eqref{5.5}.

We define the renormalized discrepancy primitive $\widetilde{\mathcal{P}}(r)$ by
\begin{equation}
\frac{d\widetilde{\mathcal{P}}}{dr}=\Delta\mathcal{D}(r)=\mathcal{D}(r)-\mathcal{D}_\text{ref}(r),
\label{5.10}
\end{equation}
together with the boundary condition that it vanish at the matching end:
\begin{equation}
\lim_{r\to\infty}\widetilde{\mathcal{P}}(r)=0,
\label{5.11}
\end{equation}
for an asymptotically flat matching regime, or
\begin{equation}
\lim_{r\to r_{\rm ref}}\widetilde{\mathcal{P}}(r)=0.
\label{5.12}
\end{equation}
We henceforth denote this unique, reference-normalized primitive by
\begin{equation}
\mathcal{P}_e(r)\equiv \widetilde{\mathcal{P}}(r).
\label{eq:Pe_is_Ptilde}
\end{equation}

Under the admissibility conditions \eqref{5.6}-\eqref{5.7}, Eqs. \eqref{5.10}-\eqref{5.12} give the explicit representations
\begin{equation}
\widetilde{\mathcal{P}}(r)=-\int_{r}^{\infty}\Delta\mathcal{D}(u)\,du,
\label{5.13}
\end{equation}
and
\begin{equation}
\widetilde{\mathcal{P}}(r)=-\int_{r}^{r_{\rm ref}}\Delta\mathcal{D}(u)\,du.
\label{5.14}
\end{equation}
The sign is chosen so that \eqref{5.11}-\eqref{5.12} hold manifestly. Differentiating \eqref{5.13} or \eqref{5.14} reproduces \eqref{5.10} whenever the improper integrals converge.

It is sometimes useful to relate $\widetilde{\mathcal{P}}$ to arbitrary representatives of the unrenormalized primitives. For any $\mathcal{P}$ and $\mathcal{P}_\text{ref}$ satisfying \eqref{5.8}-\eqref{5.9}, the difference $\mathcal{P}-\mathcal{P}_\text{ref}$ satisfies \eqref{5.10} but retains an additive ambiguity. The boundary condition \eqref{5.11} or \eqref{5.12} selects the unique representative
\begin{equation}
\widetilde{\mathcal{P}}(r)=\mathcal{P}(r)-\mathcal{P}_\text{ref}(r)-\mathcal{C},
\label{5.15}
\end{equation}
with $\mathcal{C}$ fixed so that $\widetilde{\mathcal{P}}$ vanishes in the reference limit. In particular, for a Minkowski reference on the equatorial optical manifold, $K_\text{ref}=0$ and one may take $\mathcal{P}_\text{ref}=0$, so $\widetilde{\mathcal{P}}$ is the unique representative of $\mathcal{P}$ that vanishes asymptotically.

This normalization does not restore invariance of the curvature-area term under $\mathcal{P}\mapsto\mathcal{P}+{\rm const}$, which is already built into Eq. \eqref{4.5}. Its role is to provide a canonical, endpoint-anchored representative suited to taking the closure curve into the matching end without introducing orbit-based conditions. The resulting primitive $\mathcal{P}_e$ is fixed by the chosen reference geometry and the integrability conditions \eqref{5.6}-\eqref{5.7}.

\subsection{Photon-sphere-free master expression for finite-distance deflection} \label{sec5.3}
We combine Eq. \eqref{3.7} with the reference-renormalized primitive (Eqs. \eqref{5.10}-\eqref{5.14}) to obtain a master expression whose additive gauge is fixed by the reference-endpoint normalization (Eqs. \eqref{5.11}-\eqref{5.12}), without invoking any circular null orbit.

We start from Eq. \eqref{3.7},
\begin{equation}
\alpha=- \iint_{D} K\, dS \;+\; \int_{C_\Gamma}\kappa_g\, d\ell \;+\; \phi_{RS}.
\label{5.16}
\end{equation}
where $D$ is the finite domain on the optical manifold introduced in Section \ref{sec3}. As in Section \ref{sec4}, we assume the ray can be written as a single-valued function $r=r(\phi)$ on $\phi\in[\phi_S,\phi_R]$ and choose a closure curve at $r=r_\Gamma$ in the matching regime with $r_\Gamma \ge \max_\phi r(\phi)$. Then Eq. \eqref{4.2} can be written compactly as
\begin{equation}
\iint_D K\,dS =\int_{r=r(\phi)} \mathcal{D}(r)\,d\phi,
\label{5.17}
\end{equation}
where we introduce the shorthand
\[
\int_{r=r(\phi)}\mathcal{D}(r)\,d\phi \equiv \int_{\phi_S}^{\phi_R}\left(\int_{r(\phi)}^{r_\Gamma}\mathcal{D}(r)\,dr\right)d\phi,
\]
and with $\mathcal{D}(r)=K(r)\sqrt{\det\bar{g}(r)}$ as in Eq. \eqref{4.1}.
Split the curvature density using Eq. \eqref{5.5}:
\begin{equation}
\mathcal{D}(r)=\mathcal{D}_\text{ref}(r)+\Delta\mathcal{D}(r),
\label{5.18}
\end{equation}
where $\mathcal{D}_\text{ref}(r)=K_\text{ref}(r)\sqrt{\det\bar g_\text{ref}(r)}$ and
$\Delta\mathcal{D}(r)\equiv \mathcal{D}(r)-\mathcal{D}_\text{ref}(r)$.
Substituting Eq. \eqref{5.18} into Eq. \eqref{5.17} gives the corresponding decomposition of the curvature-area integral,
\begin{equation}
\iint_{D} K\, dS=
\int_{r=r(\phi)} \mathcal{D}_\text{ref}(r)\,d\phi
+\int_{r=r(\phi)} \Delta\mathcal{D}(r)\,d\phi.
\label{5.19}
\end{equation}

To evaluate the discrepancy term in Eq. \eqref{5.19} without orbit normalization (e.g.\ fixing the additive constant by
$\mathcal{P}(r_{\rm co})=0$ at a circular null orbit), we use the renormalized discrepancy primitive $\mathcal{P}_e(r)$ defined
by Eq. \eqref{5.10} together with the endpoint normalization \eqref{5.11} or \eqref{5.12}. Applying the primitive identity
\eqref{4.4} to the radial strip integral yields
\begin{equation}
\int_{r=r(\phi)} \Delta\mathcal{D}(r)\,d\phi
=
\int_{\phi_S}^{\phi_R}\left[\mathcal{P}_e(r_\Gamma)-\mathcal{P}_e(r(\phi))\right]\,d\phi.
\label{5.20}
\end{equation}
We send the closure curve to the matching end: $r_\Gamma\to\infty$ (asymptotically flat) or $r_\Gamma\to r_{\rm ref}$ in a
finite static patch. By the endpoint normalization, this enforces
\begin{equation}
\mathcal{P}_e(r_\Gamma)\to 0
\qquad\text{as}\qquad
r_\Gamma\to \infty\ \ \text{or}\ \ r_\Gamma\to r_{\rm ref},
\label{5.21}
\end{equation}
and hence the discrepancy contribution becomes
\begin{align}
\int_{r=r(\phi)} \Delta\mathcal{D}(r)\,d\phi
&=
-\int_{\phi_S}^{\phi_R}\mathcal{P}_e(r(\phi))\,d\phi \nonumber \\
&=
-\int_{\phi_S}^{\phi_R}\widetilde{\mathcal{P}}(r(\phi))\,d\phi,
\label{5.22}
\end{align}
where we used $\mathcal{P}_e\equiv\widetilde{\mathcal{P}}$ (Eq. \eqref{eq:Pe_is_Ptilde}).
This is the key simplification: the deviation from the reference is encoded by the endpoint-normalized primitive evaluated
along the ray, with no orbit-normalization input.

For the reference term in Eq. \eqref{5.19}, define $\mathcal{P}_\text{ref}^{\star}(r)$ by
\begin{equation}
\left(\mathcal{P}_\text{ref}^{\star}\right)'(r)=\mathcal{D}_\text{ref}(r),
\label{5.23}
\end{equation}
with the reference-endpoint normalization
\begin{equation}
\mathcal{P}_\text{ref}^{\star}(r_{\rm ref})=0.
\label{5.24}
\end{equation}
Equivalently, one may write
\begin{equation}
\mathcal{P}_\text{ref}^{\star}(r)=-\int_{r}^{r_{\rm ref}}\mathcal{D}_\text{ref}(u)\,du,
\label{eq:Prefstar_integral}
\end{equation}
Under the admissibility conditions \eqref{5.6}-\eqref{5.7}, the analogue of Eqs. \eqref{5.13}-\eqref{5.14} shows that such a representative exists whenever the reference geometry is fixed. In particular, applying Eq. \eqref{4.4} to the reference strip integral gives
\begin{equation}
\int_{r=r(\phi)} \mathcal{D}_\text{ref}(r)\,d\phi
=
\int_{\phi_S}^{\phi_R}\left[\mathcal{P}_\text{ref}^{\star}(r_\Gamma)-\mathcal{P}_\text{ref}^{\star}(r(\phi))\right]\,d\phi.
\label{5.25}
\end{equation}
Taking $r_\Gamma\to r_{\rm ref}$ and using the reference-endpoint normalization \eqref{5.24} yields
\begin{equation}
\mathcal{P}_\text{ref}^{\star}(r_\Gamma)\to 0
\qquad\text{as}\qquad r_\Gamma\to r_{\rm ref},
\label{5.26}
\end{equation}
so that
\begin{equation}
\int_{r=r(\phi)} \mathcal{D}_\text{ref}(r)\,d\phi
=
-\int_{\phi_S}^{\phi_R}\mathcal{P}_\text{ref}^{\star}(r(\phi))\,d\phi.
\label{5.27}
\end{equation}
Combining Eqs. \eqref{5.19}, \eqref{5.22}, and \eqref{5.27} gives
\begin{equation}
\iint_{D} K\, dS=
-\int_{\phi_S}^{\phi_R}\left[\mathcal{P}_\text{ref}^{\star}(r(\phi))+\widetilde{\mathcal{P}}(r(\phi))\right]\,d\phi.
\label{5.28}
\end{equation}
Equation \eqref{5.28} is the photon-sphere-free master primitive reduction for finite-distance deflection.

In a finite static patch, $r_{\rm ref}$ fixes the additive gauge. If one chooses a different admissible $r_{\rm ref}'$ in $\mathcal{I}_\text{match}$, then
\begin{align}
\mathcal{P}_e(r,r_{\rm ref}')&=\mathcal{P}_e(r,r_{\rm ref})+c, \quad c=\mathcal{P}_e(r_{\rm ref},r_{\rm ref}'), 
\label{eq:rref_shift_Pe} \\
\mathcal{P}_\text{ref}^{\star}(r,r_{\rm ref}')&=\mathcal{P}_\text{ref}^{\star}(r,r_{\rm ref})+\tilde{c}, \quad \tilde{c}=\mathcal{P}_\text{ref}^{\star}(r_{\rm ref},r_{\rm ref}'),
\label{eq:rref_shift_Pref}
\end{align}
and one finds
\begin{equation}
\mathcal{P}_e(r,r_{\rm ref}')+\mathcal{P}_\text{ref}^{\star}(r,r_{\rm ref}')
=\mathcal{P}_e(r,r_{\rm ref})+\mathcal{P}_\text{ref}^{\star}(r,r_{\rm ref})+c+\tilde{c}.
\label{eq:rref_shift_sum}
\end{equation}
If one simultaneously changes the outer closure curve from $C_\Gamma(r_{\rm ref})$ to $C_\Gamma(r_{\rm ref}')$, then the difference of the corresponding boundary terms is
\begin{equation}
\int_{C_\Gamma(r_{\rm ref}')}\kappa_g\,d\ell-\int_{C_\Gamma(r_{\rm ref})}\kappa_g\,d\ell = -c-\tilde{c},
\label{eq:rref_shift_boundary}
\end{equation}
because the annular region between the two closure arcs is bounded by $\phi=\text{const}$ geodesics. Therefore the full deflection angle computed from Eq. \eqref{3.7} is independent of $r_{\rm ref}$ when the corresponding closure curve is adjusted consistently. Any residual dependence can arise only after a weak-field truncation of the exact expressions.

Two immediate consistency checks follow directly.
First, in asymptotically flat spacetimes with Minkowski reference, one has $D_{\rm ref}=0$ and therefore $P_{\rm ref}^{\star}=0$. For the Ishihara closure curve at $r_\Gamma\to\infty$, $\kappa_g\,d\ell \to d\phi$ along $C_\Gamma$, so that $\int_{C_\Gamma}\kappa_g\,d\ell = \phi_S-\phi_R=-\phi_{RS}$. Equation \eqref{5.28} simplifies Eq. \eqref{5.16} to
\begin{equation}
\alpha=
\int_{\phi_S}^{\phi_R}\widetilde{\mathcal{P}}(r(\phi))\,d\phi,
\label{5.29}
\end{equation}
Second, in the \textit{no-lens} limit where physical geometry coincides with the reference geometry, $\Delta\mathcal{D}=0$ and hence $\mathcal{P}_e\equiv 0$. Then Eq. \eqref{5.28} reduces to the reference Gauss-Bonnet identity
\begin{equation}
\alpha=
- \iint_{D} K_{\rm ref}\, dS \;+\; \int_{C_\Gamma}\kappa_g\, d\ell \;+\; \phi_{RS}.
\label{5.30}
\end{equation}
In particular, for Minkowski reference $K_{\rm ref}=0$ so $P_{\rm ref}^{\star}=0$, while $\int_{C_\Gamma}\kappa_g\,d\ell=-\phi_{RS}$, and thus $\alpha=0$ identically. This is the operational no-lens limit and motivates the endpoint normalization in Eq. \eqref{5.24}.

\subsection{Compatibility with the orbit-normalized prescription when applicable} \label{sec5.4}
The boundary-difference structure of the primitive reduction is invariant under the additive gauge transformation $\mathcal{P}\to \mathcal{P}+C$:
\begin{align}
\label{eq:invariance_under_constant_shift}
\int_{\phi_S}^{\phi_R}\!&\big[\mathcal{P}(r_{\rm ref})-\mathcal{P}(r(\phi))\big]\,d\phi= \nonumber \\
\int_{\phi_S}^{\phi_R}\!&\big[(\mathcal{P}+C)(r_{\rm ref})-(\mathcal{P}+C)(r(\phi))\big]\,d\phi.
\end{align}

Let $\mathcal{P}_\text{orb}(r)$ be the orbit-normalized primitive of the physical curvature density,
\begin{equation}
\frac{d\mathcal{P}_\text{orb}}{dr}=\mathcal{D}(r),
\label{5.31}
\end{equation}
together with the orbit-normalization condition
\begin{equation}
\mathcal{P}_\text{orb}(r_\text{co})=0,
\label{5.32}
\end{equation}
where $r_\text{co}$ is the radius of a circular null orbit (when one exists in the domain where the formalism is applicable).
Independently, let $\mathcal{P}^{\star}(r)$ be a reference-normalized primitive of the same curvature density, defined as a solution of the same first-order equation and fixed by the reference-endpoint condition
\begin{equation}
\mathcal{P}^{\star}(r)\to 0\quad\text{as}\quad r\to \infty\ \ \text{or}\ \ r\to r_{\rm ref}.
\label{5.34}
\end{equation}
Such a $\mathcal{P}^{\star}$ exists precisely under the admissible matching conditions of Section \ref{sec5.1}.

Since both $\mathcal{P}_\text{orb}$ and $\mathcal{P}^{\star}$ solve the same differential equation for $\mathcal{D}(r)$, they differ only by a constant. Evaluating at $r=r_{\rm ref}$ gives
\begin{equation}
\mathcal{P}^{\star}(r)=\mathcal{P}_\text{orb}(r)-\mathcal{P}_\text{orb}(r_{\rm ref}),
\label{5.35}
\end{equation}
where $r_{\rm ref}$ denotes the reference endpoint (either $\infty$ or a finite static-patch boundary).

We now compare the primitive reductions of the curvature-area integral. When the outer closure is taken to $r=r_{\rm ref}$, the orbit-normalized primitive gives
\begin{equation}
\iint_D K\,dS=
\int_{\phi_S}^{\phi_R}\left[\mathcal{P}_\text{orb}(r_{\rm ref})-\mathcal{P}_\text{orb}(r(\phi))\right]\,d\phi,
\label{5.36}
\end{equation}
On the other hand, the reference-normalized primitive gives
\begin{align}
\iint_D K\,dS&=
\int_{\phi_S}^{\phi_R}\left[\mathcal{P}^{\star}(r_{\rm ref})-\mathcal{P}^{\star}(r(\phi))\right]\,d\phi \nonumber \\
&=-\int_{\phi_S}^{\phi_R}\mathcal{P}^{\star}(r(\phi))\,d\phi,
\label{5.37}
\end{align}
where the last equality uses Eq. \eqref{5.34}. Substituting Eq. \eqref{5.35} into Eq. \eqref{5.37} yields
\begin{align}
-\int_{\phi_S}^{\phi_R}\mathcal{P}^{\star}(r(\phi))\,&d\phi=
-\int_{\phi_S}^{\phi_R}\left[\mathcal{P}_\text{orb}(r(\phi))-\mathcal{P}_\text{orb}(r_{\rm ref})\right]\,d\phi \nonumber \\ &=
\int_{\phi_S}^{\phi_R}\left[\mathcal{P}_\text{orb}(r_{\rm ref})-\mathcal{P}_\text{orb}(r(\phi))\right]\,d\phi,
\label{5.38}
\end{align}
which is exactly Eq. \eqref{5.36}. Therefore the curvature-area contribution is identical whether one uses orbit normalization or reference normalization. Since $\phi{RS}$ is the same geometric quantity in Eq. \eqref{3.7} and Eq. \eqref{2.11}, the full deflection angle agrees in the overlap regime where both normalizations are available, while the present formalism remains meaningful when no circular null orbit is available.

Finally, we note how this compatibility appears at the level of the discrepancy primitive. In asymptotically flat cases with Minkowski reference, $\mathcal{P}_\text{ref}^{\star}=0$ and Eq. \eqref{5.28} reduces to Eq. \eqref{5.29}. If, in addition, an orbit-normalized primitive $\mathcal{P}_\text{orb}$ is introduced, then Eq. \eqref{5.35} shows that $\widetilde{\mathcal{P}}$ coincides with the reference-normalized physical primitive $\mathcal{P}^{\star}$, which in turn differs from $\mathcal{P}_\text{orb}$ only by the constant $\mathcal{P}_\text{orb}(\infty)$. Thus the method selects the same primitive class but anchors it at the reference endpoint, consistent with the operational definition of $\alpha$.

\section{General properties and domain of validity} \label{sec6}
We record general features of the reference-renormalized formulation of Section \ref{sec5} and state the conditions under which the finite-distance Gauss-Bonnet construction and the primitive reduction are mathematically and operationally sound. We emphasize coordinate robustness within the static spherical class and delimit the weak-lensing regime in which the observable in Eq. \eqref{2.11} is unambiguous. In this sense, $\alpha$ is a geometric invariant of the chosen configuration.

\subsection{Coordinate robustness within the spherical class} \label{sec6.1}
The deflection angle $\alpha$ is defined operationally by local angles and the endpoint separation in Eq. \eqref{2.11}; Eq. \eqref{3.7} rewrites the same observable using intrinsic curvature on the optical manifold plus the same endpoint separation. Any coordinate dependence can therefore enter only through intermediate representations, such as writing the ray as $r=r(\phi)$ or choosing a radial coordinate to express the optical metric.
Consider a monotonic radial reparameterization $r\mapsto \bar r(r)$ with $dr/d\bar r>0$ on the region of interest. The spacetime line element in Eq. \eqref{2.1} retains its static spherical form after this change, with transformed metric functions $\bar A(\bar r)=A(r(\bar r))$, $\bar C(\bar r)=C(r(\bar r))$, and $\bar B(\bar r)=B(r(\bar r))\,(dr/d\bar r)^2$. The equatorial optical metric in Eq. \eqref{2.6} then transforms covariantly as
\begin{equation}
d\sigma^{2}
=\frac{B(r)}{A(r)}\,dr^{2}+\frac{C(r)}{A(r)}\,d\phi^{2}
=\frac{\bar B(\bar r)}{\bar A(\bar r)}\,d\bar r^{2}+\frac{\bar C(\bar r)}{\bar A(\bar r)}\,d\phi^{2}.
\label{6.1}
\end{equation}
Since Gaussian curvature $K$ is an intrinsic scalar of the two-metric and $dS=\sqrt{\det\bar g}\,dr\,d\phi$ is its associated area element, the curvature two-form $K\,dS$ is invariant. Therefore the domain integral in Eq. \eqref{3.7} is independent of the chosen radial coordinate provided the domain $D$ is specified geometrically on the optical manifold.

The primitive reduction appears coordinate-dependent because the primitive is defined through differentiation with respect to a chosen radial variable. In the physical geometry we introduced the curvature density $\mathcal{D}(r)=K(r)\sqrt{\det\bar g(r)}$ and a primitive satisfying $d\mathcal{P}/dr=\mathcal{D}(r)$, while in the reference-renormalized formalism we use the discrepancy density $\Delta\mathcal{D}(r)$ and the renormalized primitive satisfying Eq. \eqref{5.10}. 

Under $r\mapsto \bar r$, the invariant curvature 2-form
$\Delta K\,dS=\Delta K(r)\sqrt{\det\bar g(r)}\,dr\,d\phi$
is unchanged. Equivalently, the radial curvature density
$\Delta\mathcal{D}(r)\equiv \Delta K(r)\sqrt{\det\bar g(r)}$
transforms with the Jacobian so that
$\Delta\mathcal{D}(r)\,dr=\Delta\bar{\mathcal{D}}(\bar r)\,d\bar r$.
Therefore Eq. \eqref{5.10} may be written in either coordinate as
\begin{equation}
\frac{d\widetilde{\mathcal{P}}}{dr}=\Delta\mathcal{D}(r)
\qquad\Longleftrightarrow\qquad
\frac{d\widetilde{\mathcal{P}}}{d\bar r}=\Delta\bar{\mathcal{D}}(\bar r),
\label{6.2}
\end{equation}
Once the endpoint normalization \eqref{5.11} or \eqref{5.12} is imposed, $\widetilde{\mathcal{P}}$ is a scalar function on the optical manifold and its differences are coordinate-independent:
\begin{equation}
\widetilde{\mathcal{P}}(r_2)-\widetilde{\mathcal{P}}(r_1)
=\int_{r_1}^{r_2}\Delta\mathcal{D}(u)\,du
=\int_{\bar r_1}^{\bar r_2}\Delta\bar{\mathcal{D}}(v)\,dv .
\end{equation}
The master expression, such as Eq. \eqref{5.28} or its asymptotically flat specialization Eq. \eqref{5.29}, depends only on evaluations of $\widetilde{\mathcal{P}}$ along the ray composed with $r(\phi)$ and on the coordinate separation $\phi_{RS}$. Since $\phi$ is fixed by the rotational Killing symmetry up to an additive constant, $\phi_{RS}$ is invariant under $\phi\mapsto\phi+\mathrm{const}$, and $\int_{\phi_S}^{\phi_R}\widetilde{\mathcal{P}}(r(\phi))\,d\phi$ is likewise invariant. Therefore $\phi_{RS}$ is unaffected by shifting the origin of $\phi$.

Writing the ray as $r=r(\phi)$ is a convenience valid when the projected trajectory is simple (no self-intersection or caustics) in the weak-deflection regime. If instead one parameterizes the ray by an affine parameter or by the optical arclength, the curvature and boundary terms take a different form, but the combined observable in Eq. \eqref{3.7} is unchanged.

\subsection{Weak-deflection regime and finite-distance constraints} \label{sec6.2}
The Gauss-Bonnet identity in Eq. \eqref{3.1} is exact, but its lensing interpretation requires a domain $D$ on a Riemannian optical manifold and a ray that represents a single bending event. The primitive reduction likewise assumes the ray can be treated as single-valued in $\phi$ over the endpoint interval. We summarize the conditions.

First, the optical metric in Eq. \eqref{2.6} is Riemannian only in the static region where
$A(r)>0$, $B(r)>0$, and $C(r)>0$ (equivalently $\bar g_{rr}>0$ and $\bar g_{\phi\phi}>0$). We therefore restrict to configurations where the full Gauss-Bonnet domain $D$ lies within such a region. This excludes domains that cross horizons in black-hole spacetimes or that approach coordinate singularities where the optical metric degenerates. In practice, for black-hole metrics we place the source and receiver at radii larger than the relevant horizon radius and restrict to bending trajectories whose turning point, when present, also lies outside the horizon. For Kottler-type geometries, we likewise restrict to the static patch bounded by the cosmological horizon and choose the reference endpoint in Eq. \eqref{5.12} accordingly. In the Kottler example (Sec. \ref{8.3}) we further assume that the endpoints and the turning point (when present) lie well
inside the static patch so that $\Lambda r^2\ll 1$ for the radii involved, and we work in the weak-deflection regime
$r_g/b\ll 1$ and $\Lambda b^2\ll 1$. These conditions ensure that the de Sitter reference and the matching regime are
simultaneously available and that the endpoint angles are operationally defined for static observers.

Second, the endpoint angles $\Psi_S$ and $\Psi_R$ defined in Eq. \eqref{2.10} require that $\sin\Psi(r)=b\sqrt{A(r)/C(r)}$ lie in the interval $[0,1]$ at the endpoints. This implies the kinematic constraint
\begin{equation}
b^{2}\leq \frac{C(r_S)}{A(r_S)},\qquad b^{2}\leq \frac{C(r_R)}{A(r_R)},
\label{6.3}
\end{equation}
which ensures that the local direction of the ray is physically realizable at both $S$ and $R$. In weak lensing one typically has $b$ much smaller than the endpoint radii in the relevant reference sense, so Eq. \eqref{6.3} is automatically satisfied, but at finite distance it is essential to keep the inequality explicit.
Third, weak deflection assumes the ray does not wind around the center and that $\phi_{RS}$ stays close to the reference value. Then the projected ray segment between $S$ and $R$ is a simple curve and the domain $D$ can be chosen to be simply connected with Euler characteristic $\chi(D)=1$. It also justifies representing the ray as a single-valued function $r=r(\phi)$ on the interval $[\phi_S,\phi_R]$. When strong deflection occurs, multiple images and multiple windings arise; in that situation the construction must be modified to account for self-intersecting boundaries and possibly multiple domains, and the primitive reduction in its simplest form is no longer sufficient \cite{Ishihara:2016sfv,Bozza:2002zj}.

Fourth, the reference-renormalized construction requires admissible matching conditions for the discrepancy density (Eqs. \eqref{5.6}-\eqref{5.7}) and, when $\mathcal{P}^{\star}_{\rm ref}$ is used explicitly as in Eq. \eqref{5.28}, the corresponding reference integrability condition (Eq. \eqref{V.ref_adm}). Physically, this is the requirement that there exists a regime in which the physical optical geometry approaches the chosen reference optical geometry sufficiently well that the discrepancy curvature density is integrable toward the reference endpoint. This is the minimal condition ensuring that $\widetilde{\mathcal{P}}$ is uniquely defined by Eqs. \eqref{5.11}-\eqref{5.12} and that the limit $r_\Gamma\to r_{\rm ref}$ used in Eq. \eqref{5.21} is legitimate. In asymptotically flat cases this is realized at large radius with Minkowski reference, while in Kottler-type cases it is realized in the static patch with de Sitter reference.
Taken together, these conditions define the intended scope of the method: finite-distance, single-pass weak lensing by static spherical geometries in a regular optical region, with a well-defined reference regime.

\subsection{Limiting-case checks} \label{sec6.3}
We record structural checks that will be used implicitly when specializing to concrete metrics.

The first check is the reference limit. If the physical metric coincides with the chosen reference metric, then $\Delta\mathcal{D}(r)=0$ and Eq. \eqref{5.10} implies $\widetilde{\mathcal{P}}(r)$ is constant. The normalization condition in Eq. \eqref{5.11} or Eq. \eqref{5.12} then forces $\widetilde{\mathcal{P}}(r)=0$ identically. In this situation the master expression in Eq. \eqref{5.28} reduces to the primitive form of the Gauss-Bonnet identity for the reference optical manifold. Since $\alpha$ is defined by local angles and endpoint separation in Eq. \eqref{2.11}, no lens corresponds precisely to the statement that $\Psi_R-\Psi_S+\phi_{RS}=0$, and reference normalization enforces this without invoking any special orbit.

The second check is the large-distance limit when the matching endpoint is spatial infinity. For fixed $b$, the endpoint angles in Eq. \eqref{2.10} satisfy $\Psi(r)\to 0$ as $r\to\infty$ in asymptotically flat geometries with $A(r)\to 1$ and $C(r)\sim r^{2}$. The finite-distance definition in Eq. \eqref{2.11} then reduces to the usual asymptotic comparison between incoming and outgoing directions, and Eq. \eqref{3.7} becomes the standard Gauss-Bonnet representation of asymptotic bending. The reference-renormalized primitive is tailored to this limit because Eq. \eqref{5.11} enforces $\widetilde{\mathcal{P}}(\infty)=0$, ensuring that no spurious constant survives from the primitive reduction as the outer closure is taken to infinity.

The third check concerns endpoint dependence. At finite $r_S$ and $r_R$, Eq. \eqref{2.10} shows that $\alpha$ necessarily contains explicit finite-distance corrections through $\Psi_S$ and $\Psi_R$. The Gauss-Bonnet representation organizes these corrections through the geometry of $D$, but it does not eliminate them. In later weak-field expansions, endpoint dependence appears both through the local angles and through the evaluation of the primitive along a trajectory determined by $b$, $r_S$, and $r_R$.

Finally, orbit normalization is recovered as a special case. When a circular null orbit exists in the optical region and the orbit-normalized primitive of Eqs. \eqref{5.31}-\eqref{5.32} is well defined, Section \ref{sec5.4} shows that the curvature contribution, and therefore $\alpha$, agrees exactly with the reference-normalized construction. This compatibility follows from the constant-shift equivalence of primitives in Eq. \eqref{5.35} and the identity of curvature contributions in Eq. \eqref{5.38}.

\section{Practical evaluation strategy for weak deflection without orbit-normalization input} \label{sec7}
We outline a weak-deflection implementation of the reference-renormalized primitive introduced in Section \ref{sec5} for finite source and receiver distances. The inputs are endpoint radii $(r_S,r_R)$, the impact parameter $b$, the metric functions and their reference counterparts, and one-dimensional integrals whose expansions can be organized systematically. No step requires the existence of a circular null orbit.

Here \textit{without orbit input} refers only to the \emph{primitive normalization} step: we do not fix the additive constant by imposing a vanishing condition at a privileged orbit (e.g. a circular null orbit/photon sphere), and we do not introduce an externally chosen orbit radius as a lower limit for the primitive. Finite-distance lensing still depends on null-geodesic kinematics through $\phi_{RS}$ and through the evaluation of primitives along the physical ray; this dependence enters through the orbit relation \eqref{7.1} and the turning-point condition \eqref{7.2}, with $r_0$ determined implicitly by $b$ and endpoint data, rather than supplied as an independent normalization input.

\subsection{Parameterization of the ray and endpoint relations} \label{sec7.1}
We work with null geodesics in the equatorial plane $\theta=\pi/2$ of the static spherical spacetime (Eq. \eqref{2.1}). The constants of motion $E$ and $L$ in Eq. \eqref{2.2} define the (nonnegative) impact parameter
\begin{equation}
b\equiv \frac{|L|}{E}\ge 0,
\label{7.bdef}
\end{equation}
while the sign of $L$ fixes the orientation of $\phi$ along the ray. We adopt the convention that $\phi$ increases from the source branch through the turning point to the receiver branch, so that $L>0$ and $d\phi/d\lambda=L/C(r)>0$ on the receiver side. Using Eq. \eqref{2.4} together with $d\phi/d\lambda=L/C(r)$, we obtain
\begin{equation}
\frac{d\phi}{dr}=
\pm\,\frac{b\,\sqrt{A(r)\,B(r)}}{\sqrt{C(r)}\,\sqrt{C(r)-A(r)\,b^{2}}}.
\label{7.1}
\end{equation}
When the trajectory has a turning point, it occurs at $r=r_0$ where $dr/d\lambda=0$, equivalently where the denominator in Eq. \eqref{7.1} vanishes:
\begin{equation}
b^{2}=\frac{C(r_0)}{A(r_0)}.
\label{7.2}
\end{equation}
We set the azimuthal origin so that $\phi=0$ at the turning point. The ray then connects the source at $\phi=\phi_S<0$ to the receiver at $\phi=\phi_R>0$, and
\begin{equation}
\phi_{RS}\equiv \phi_R-\phi_S.
\label{7.3}
\end{equation}
With
\begin{equation}
\Phi(r,b)\equiv \int_{r_0}^{r}\frac{b\,\sqrt{A(u)\,B(u)}}{\sqrt{C(u)}\,\sqrt{C(u)-A(u)\,b^{2}}}\,du,
\label{7.4}
\end{equation}
we have
\begin{equation}
\phi_R=\Phi(r_R,b),\qquad \phi_S=-\Phi(r_S,b),
\label{7.5}
\end{equation}
and hence
\begin{equation}
\phi_{RS}=\Phi(r_R,b)+\Phi(r_S,b).
\label{7.6}
\end{equation}
The local endpoint angles in the observable \eqref{2.11} are fixed by Eq. \eqref{2.10}; in particular
\begin{equation}
\Psi(r)=\arcsin\!\left(b\,\sqrt{\frac{A(r)}{C(r)}}\right),
\label{7.7}
\end{equation}
so that $\Psi_S=\Psi(r_S)$ and $\Psi_R=\Psi(r_R)$. The kinematic constraint \eqref{6.3} ensures that Eq. \eqref{7.7} is well defined.

\subsection{Expansion scheme for metric functions and renormalized primitive} \label{sec7.2}
Let $(A_\text{ref},B_\text{ref},C_\text{ref})$ define the reference spacetime \eqref{5.1}. We treat the physical metric as a small deformation of the reference controlled by a bookkeeping parameter $\varepsilon$ (identified with the appropriate dimensionless combinations in concrete metrics), writing
\begin{align}
A(r)&=A_\text{ref}(r)\left[1+\varepsilon\,a_1(r)+\varepsilon^{2}a_2(r)+\mathcal{O}(\varepsilon^3)\right],\nonumber \\
B(r)&=B_\text{ref}(r)\left[1+\varepsilon\,b_1(r)+\varepsilon^{2}b_2(r)+\mathcal{O}(\varepsilon^3)\right],
\label{7.8}
\end{align}
\begin{equation}
C(r)=C_\text{ref}(r)\left[1+\varepsilon\,c_1(r)+\varepsilon^{2}c_2(r)+\mathcal{O}(\varepsilon^3)\right],
\label{7.9}
\end{equation}
The turning point determined by Eq. \eqref{7.2} is expanded as
\begin{equation}
r_0=r_0^{(0)}+\varepsilon\,r_0^{(1)}+\varepsilon^{2}r_0^{(2)}+\mathcal{O}(\varepsilon^3),
\label{7.10}
\end{equation}
with $r_0^{(0)}$ fixed by $b^{2}=C_\text{ref}(r_0^{(0)})/A_\text{ref}(r_0^{(0)})$ and higher coefficients obtained by expanding Eq. \eqref{7.2} order by order.

For the curvature contribution, expand the discrepancy density $\Delta\mathcal{D}(r)$ and the renormalized primitive,
\begin{equation}
\Delta\mathcal{D}(r)=\varepsilon\Delta\mathcal{D}_1(r)+\varepsilon^{2}\Delta\mathcal{D}_2(r)+\mathcal{O}(\varepsilon^3).
\label{7.11}
\end{equation}
\begin{equation}
\widetilde{\mathcal{P}}(r)=\varepsilon\,\widetilde{\mathcal{P}}_1(r)+\varepsilon^{2}\widetilde{\mathcal{P}}_2(r)+\mathcal{O}(\varepsilon^3),
\label{7.12}
\end{equation}
with coefficients fixed by reference-endpoint integrals:
\begin{equation}
\widetilde{\mathcal{P}}_{n}(r)=-\int{r}^{\infty}\Delta\mathcal{D}_{n}(u)\,du,
\qquad n=1,2,
\label{7.13}
\end{equation}
\begin{equation}
\widetilde{\mathcal{P}}_{n}(r)=-\int_{r}^{r_{\rm ref}}\Delta\mathcal{D}_{n}(u)\,du,
\qquad n=1,2.
\label{7.14}
\end{equation}
Equations \eqref{7.13}-\eqref{7.14} fix $\widetilde{\mathcal{P}}$ uniquely once $\Delta\mathcal{D}_n(r)$ is computed.

For explicit evaluation it is often convenient to convert $\phi$-integrals to radial integrals using Eq. \eqref{7.1}. For any sufficiently regular $F(r)$ along the ray,
\begin{align}
\int_{\phi_S}^{\phi_R}F(r(\phi))\,&d\phi=
\int_{r_0}^{r_S}F(r)\,\frac{b\,\sqrt{A(r)\,B(r)}}{\sqrt{C(r)}\,\sqrt{C(r)-A(r)\,b^{2}}}\,dr
\nonumber \\
&+
\int_{r_0}^{r_R}F(r)\,\frac{b\,\sqrt{A(r)\,B(r)}}{\sqrt{C(r)}\,\sqrt{C(r)-A(r)\,b^{2}}}\,dr,
\label{7.15}
\end{align}
Applying Eq. \eqref{7.15} to $F=\widetilde{\mathcal{P}}$ and to the reference primitive $\mathcal{P}^{\star}_\text{ref}$ in Eq. \eqref{5.28} yields a purely radial representation of the deflection angle. If $\widetilde{\mathcal{P}}$ is of order $\varepsilon$, then replacing the physical orbit in the integrand by the reference orbit changes the integral only at order $\varepsilon^{2}$, justifying evaluation of primitive contributions along the reference trajectory at leading order.

\subsection{Separation of curvature and endpoint-angle contributions} \label{sec7.3}
The operational definition of the finite-distance deflection angle is
\begin{equation}
\alpha=\Psi_R-\Psi_S+\phi_{RS},
\label{7.16}
\end{equation}
with $\Psi_S,\Psi_R$ given by Eq. \eqref{7.7} and $\phi_{RS}$ by Eq. \eqref{7.6}. The Gauss-Bonnet representation \eqref{3.7} reorganizes the same observable through an intrinsic-curvature integral; in the reference-renormalized formulation this is encoded in the master relation \eqref{5.28}. Using Eq. \eqref{7.15} to convert the primitive term to radial integrals yields
\begin{align}
&\alpha=
\phi_{RS}
-\int_{r_0}^{r_S}\left[\mathcal{P}_\text{ref}^{\star}(r)+\widetilde{\mathcal{P}}(r)\right]\,
\nonumber \\
&\times \frac{b\,\sqrt{A(r)\,B(r)}}{\sqrt{C(r)}\,\sqrt{C(r)-A(r)\,b^{2}}}\,dr
-\int_{r_0}^{r_R}\left[\mathcal{P}_\text{ref}^{\star}(r)+\widetilde{\mathcal{P}}(r)\right]\,\nonumber \\
&\times\frac{b\,\sqrt{A(r)\,B(r)}}{\sqrt{C(r)}\,\sqrt{C(r)-A(r)\,b^{2}}}\,dr.
\label{7.17}
\end{align}
Here, \(\mathcal{P}_\text{ref}^{\star}\) in Eqs. \eqref{5.23}-\eqref{5.24}. Equation \eqref{7.17} is the practical evaluation formula in the new formalism: all inputs are the metric functions, the reference choice, $(r_S,r_R)$, and $b$, with $r_0$ fixed by Eq. \eqref{7.2}.

In weak deflection, expand the endpoint angles from Eq. \eqref{7.7} using \eqref{7.8}-\eqref{7.9} at $r_S$ and $r_R$, expand $\phi_{RS}$ from \eqref{7.6} using \eqref{7.4} and \eqref{7.10}, and expand the primitive contribution in \eqref{7.17} by inserting \eqref{7.12}-\eqref{7.14} and expanding the remaining orbit factor using \eqref{7.8}-\eqref{7.9}. The net result is an expression for $\alpha(r_S,r_R,b)$ organized systematically in powers of $\varepsilon$, suitable for term-by-term comparison with the finite-distance deflection formulas in Ishihara's framework for Schwarzschild, Reissner-Nordstr"om, and Kottler spacetimes.

\section{examples and agreement with Ishihara: Three cases (and a no-photon-sphere demonstration)} \label{sec8}
In this section we validate the reference-renormalized curvature-primitive formalism by applying it to three canonical
static spherical spacetimes (Schwarzschild, Reissner-Nordstr\"om, and Kottler), where direct comparison to Ishihara et al.'s
finite-distance formulas is available, and then we add a short demonstrative example where orbit-normalization is not even
formulable because no circular null orbit exists in the physical optical region \cite{Janis:1968zz,Fisher:1948yn,Wyman:1981bd,Sau:2020xau}. For each case we (i) specify the physical and reference optical geometries, (ii) compute the renormalized primitive explicitly, (iii) evaluate the finite-distance weak-deflection angle without invoking any circular null orbit, and (iv) demonstrate agreement with the finite-distance deflection obtained from the operational endpoint-angle definition introduced in Eq. \eqref{7.16}. We begin with Schwarzschild, where the reference is Minkowski and every step can be exhibited in closed form at leading order.

\subsection{Example 1: The Schwarzschild metric} \label{sec8.1}
We take the Schwarzschild metric in standard curvature coordinates,
\begin{equation}
ds^{2}= -\left(1-\frac{2M}{r}\right)dt^{2} + \left(1-\frac{2M}{r}\right)^{-1}dr^{2} + r^{2}d\Omega^2,
\label{8.1}
\end{equation}
where $M>0$ is the mass parameter and we work in the static region $r>2M$. In the notation of Eq. \eqref{2.1}, the metric functions are
\begin{equation}
A(r)=1-\frac{2M}{r},\qquad B(r)=A(r)^{-1},\qquad C(r)=r^{2}.
\label{8.2}
\end{equation}
For an asymptotically flat geometry the natural reference is Minkowski space in the same spherical chart, corresponding to
\begin{equation}
A_\text{ref}(r)=1,\quad B_\text{ref}(r)=1,\quad C_\text{ref}(r)=r^{2}.
\label{8.3}
\end{equation}
The equatorial optical metric of the physical spacetime follows from Eq. \eqref{2.6}. Using Eq. \eqref{8.2}, we obtain
\begin{equation}
d\sigma^{2}=\frac{dr^{2}}{\left(1-\frac{2M}{r}\right)^{2}}+\frac{r^{2}}{1-\frac{2M}{r}}\,d\phi^{2}.
\label{8.4}
\end{equation}
The reference optical metric is simply the flat Euclidean metric on the equatorial plane,
\begin{equation}
d\sigma_\text{ref}^{2}=dr^{2}+r^{2}\,d\phi^{2}.
\label{8.5}
\end{equation}
Since the Gaussian curvature of Eq. \eqref{8.5} vanishes, $K_\text{ref}=0$, the reference curvature density is $\mathcal{D}_\text{ref}=0$, and therefore the discrepancy density introduced in Eq. \eqref{5.5} is simply $\Delta\mathcal{D}=\mathcal{D}$ in this case.

The primitive reduction hinges on the curvature density $\mathcal{D}(r)=K(r)\sqrt{\det\bar g(r)}$, where $\bar g$ is the optical metric in Eq. \eqref{8.4}. For diagonal, $\phi$-independent two-metrics, Eq. \eqref{3.8} shows that the curvature density is a total radial derivative. We use that structure to compute the renormalized primitive in a way that is both transparent and avoids any orbit input.
For Eq. \eqref{8.4} we read off
\begin{equation}
\bar g_{rr}=\left(1-\frac{2M}{r}\right)^{-2},\qquad
\bar g_{\phi\phi}=\frac{r^{2}}{1-\frac{2M}{r}}.
\label{8.6}
\end{equation}
Hence
\begin{equation}
\sqrt{\det\bar g}=\sqrt{\bar g_{rr}\bar g_{\phi\phi}}
=\frac{r}{\left(1-\frac{2M}{r}\right)^{3/2}}.
\label{8.7}
\end{equation}
Moreover, the only connection coefficient needed in Eq. \eqref{3.8} is
\begin{align}
\bar\Gamma^{\phi}{}_{r\phi}
&=\frac{1}{2}\frac{d}{dr}\ln\bar g_{\phi\phi}
=\frac{1}{2}\frac{d}{dr}\ln\!\left(\frac{r^{2}}{1-\frac{2M}{r}}\right) \nonumber \\
&=\frac{1}{r}-\frac{M}{r^{2}\left(1-\frac{2M}{r}\right)}.
\label{8.8}
\end{align}
Substituting Eqs. \eqref{8.6}-\eqref{8.8} into Eq. \eqref{3.8}, the curvature density becomes
\begin{equation}
\mathcal{D}(r)=K(r)\sqrt{\det\bar g(r)}=
-\frac{d}{dr}\!\left(\frac{\sqrt{\det\bar g}}{\bar g_{rr}}\bar\Gamma^{\phi}{}_{r\phi}\right).
\label{8.9}
\end{equation}
A short algebraic simplification gives
\begin{equation}
\frac{\sqrt{\det\bar g}}{\bar g_{rr}}\bar\Gamma^{\phi}{}_{r\phi}=
\sqrt{1-\frac{2M}{r}}-\frac{M}{r\sqrt{1-\frac{2M}{r}}}.
\label{8.10}
\end{equation}
Therefore Eq. \eqref{8.9} yields $\mathcal{D}(r)=-dF/dr$ with
\begin{equation}
F(r)\equiv \sqrt{1-\frac{2M}{r}}-\frac{M}{r\sqrt{1-\frac{2M}{r}}}.
\label{8.11}
\end{equation}
The reference-renormalized primitive $\widetilde{\mathcal{P}}(r)$ is defined by Eq. \eqref{5.10} with the asymptotic normalization in Eq. \eqref{5.11}. Since $\Delta\mathcal{D}=\mathcal{D}$ here, we have $d\widetilde{\mathcal{P}}/dr=\mathcal{D}(r)$ and $\widetilde{\mathcal{P}}(\infty)=0$. Using $\mathcal{D}=-F'(r)$ and $F(\infty)=1$, the unique renormalized primitive is obtained immediately:
\begin{align}
\widetilde{\mathcal{P}}(r)&=
-\int_{r}^{\infty}\mathcal{D}(u)\,du
=
-\int_{r}^{\infty}\left[-F'(u)\right]\,du \nonumber \\
&=
F(\infty)-F(r)
=
1-F(r).
\label{8.12}
\end{align}
Substituting Eq. \eqref{8.11} into Eq. \eqref{8.12} yields the closed form
\begin{equation}
\widetilde{\mathcal{P}}(r)=
1-\sqrt{1-\frac{2M}{r}}+\frac{M}{r\sqrt{1-\frac{2M}{r}}}.
\label{8.13}
\end{equation}
This expression is photon-sphere-free by construction: its normalization is fixed solely by the asymptotic reference condition $\widetilde{\mathcal{P}}(\infty)=0$.
For weak deflection we expand Eq. \eqref{8.13} in $M/r$:
\begin{equation}
\widetilde{\mathcal{P}}(r)=
\frac{2M}{r}+\mathcal{O}\!\left(\frac{M^{2}}{r^{2}}\right).
\label{8.14}
\end{equation}
The leading behavior $\widetilde{\mathcal{P}}(r)\sim 2M/r$ is the only input required to obtain the leading finite-distance bending angle.

We now evaluate the deflection angle to first order in $M/b$ with finite source and receiver radii $(r_S,r_R)$. In the reference-renormalized approach, the deflection is defined relative to the chosen reference geometry. For the present asymptotically flat case with Minkowski reference, the zeroth-order (reference) trajectory is a straight line in the Euclidean optical plane with minimal distance $b$ from the origin. Choosing the closest-approach point as $\phi=0$, the straight-line polar equation is
\begin{equation}
r^{(0)}(\phi)=\frac{b}{\cos\phi},
\qquad -\phi_S^{(0)}\le \phi\le \phi_R^{(0)},
\label{8.15}
\end{equation}
with endpoint angles fixed by the Euclidean relations
\begin{align}
\cos\phi_R^{(0)}&=\frac{b}{r_R},\quad
\cos\phi_S^{(0)}=\frac{b}{r_S},
\nonumber \\
\phi_S^{(0)}&\in\left(0,\frac{\pi}{2}\right),\quad \phi_R^{(0)}\in\left(0,\frac{\pi}{2}\right),
\label{8.16}
\end{align}
so that the source point lies at $\phi=-\phi_S^{(0)}$ and the receiver at $\phi=+\phi_R^{(0)}$. The corresponding sines are
\begin{equation}
\sin\phi_R^{(0)}=\sqrt{1-\frac{b^{2}}{r_R^{2}}},\qquad
\sin\phi_S^{(0)}=\sqrt{1-\frac{b^{2}}{r_S^{2}}}.
\label{8.17}
\end{equation}
At first order in $M/b$, the curvature-discrepancy contribution may be evaluated along the reference trajectory, because replacing $r(\phi)$ by $r^{(0)}(\phi)$ in an $\mathcal{O}(M)$ integrand produces only $\mathcal{O}(M^{2})$ changes. More explicitly, along the weak-field trajectory one has $\widetilde{\mathcal P}=\mathcal{O}(M/r)=\mathcal{O}(M/b)$,
while the orbit perturbation satisfies $r(\phi)=r^{(0)}(\phi)+\mathcal{O}(M)$; hence
$\widetilde{\mathcal P}(r(\phi))-\widetilde{\mathcal P}(r^{(0)}(\phi))=\mathcal{O}(M^2/b^2)$ and the replacement is consistent at the stated order. Using Eq. \eqref{8.14}, we therefore obtain the leading bending angle as an integral of the leading renormalized primitive along the straight-line path,
\begin{align}
\alpha&=
\int_{-\phi_S^{(0)}}^{\phi_R^{(0)}} \widetilde{\mathcal{P}}(r^{(0)}(\phi))\,d\phi
+\mathcal{O}\!\left(\frac{M^{2}}{b^{2}}\right)\nonumber \\
&=
\frac{2M}{b}\int_{-\phi_S^{(0)}}^{\phi_R^{(0)}}\cos\phi\,d\phi
+\mathcal{O}\!\left(\frac{M^{2}}{b^{2}}\right).
\label{8.18}
\end{align}
Evaluating the elementary integral gives
\begin{align}
\alpha&=
\frac{2M}{b}\left[\sin\phi\right]_{-\phi_S^{(0)}}^{\phi_R^{(0)}}
+\mathcal{O}\!\left(\frac{M^{2}}{b^{2}}\right) \nonumber \\
&=
\frac{2M}{b}\left(\sin\phi_R^{(0)}+\sin\phi_S^{(0)}\right)
+\mathcal{O}\!\left(\frac{M^{2}}{b^{2}}\right).
\label{8.19}
\end{align}
Substituting Eq. \eqref{8.17} into Eq. \eqref{8.19}, we obtain the explicit finite-distance weak-deflection formula
\begin{equation}
\alpha=
\frac{2M}{b}\left(\sqrt{1-\frac{b^{2}}{r_R^{2}}}+\sqrt{1-\frac{b^{2}}{r_S^{2}}}\right)
+\mathcal{O}\!\left(\frac{M^{2}}{b^{2}}\right).
\label{8.20}
\end{equation}
This is the desired photon-sphere-free result computed solely from the renormalized primitive and the reference straight-line geometry.
Two immediate checks follow. First, in the large-distance limit $r_S,r_R\to\infty$, the square roots tend to unity and Eq. \eqref{8.20} reduces to the standard leading Schwarzschild deflection $4M/b$. Second, Eq. \eqref{8.20} respects the finite-distance kinematic constraint $b\le r_S$ and $b\le r_R$ inherited from Eq. \eqref{6.3}: the square roots are real precisely in the physically admissible regime.

To demonstrate agreement with Ishihara's finite-distance observable, we now recover Eq. \eqref{8.20} directly from the endpoint-angle definition (Eq. \eqref{7.16}) specialized to Schwarzschild, without using any curvature primitive. This provides a stringent internal consistency check: the primitive computation and the endpoint-angle computation must yield the same $\alpha$ to the order considered.
The operational definition is
\begin{equation}
\alpha=\Psi_R-\Psi_S+\phi_{RS},
\label{8.21}
\end{equation}
with $\sin\Psi=b\sqrt{A/C}$ from Eq. \eqref{2.10} and $\phi_{RS}$ obtained from the orbit equation (Eq. \eqref{7.1}). For Schwarzschild, Eq. \eqref{2.10} gives
\begin{equation}
\sin\Psi(r)=\frac{b}{r}\sqrt{1-\frac{2M}{r}}.
\label{8.22}
\end{equation}
Because the photon propagates inward at the source and outward at the receiver, the physical branch choices are
\begin{align}
\Psi_R&=\arcsin\!\left(\frac{b}{r_R}\sqrt{1-\frac{2M}{r_R}}\right)\in\left(0,\frac{\pi}{2}\right),
\nonumber \\
\Psi_S&=\pi-\arcsin\!\left(\frac{b}{r_S}\sqrt{1-\frac{2M}{r_S}}\right)\in\left(\frac{\pi}{2},\pi\right).
\label{8.23}
\end{align}
This branch structure is essential: it guarantees that in the Minkowski limit $M\to 0$ the combination in Eq. \eqref{8.21} vanishes for a straight line, as required.
We now expand Eq. \eqref{8.23} to first order in $M$. Define the convenient abbreviations
\begin{align}
x_R\equiv \frac{b}{r_R},\quad &x_S\equiv \frac{b}{r_S},\quad
F_R\equiv \sqrt{1-x_R^{2}}=\sqrt{1-\frac{b^{2}}{r_R^{2}}},\nonumber \\
&F_S\equiv \sqrt{1-x_S^{2}}=\sqrt{1-\frac{b^{2}}{r_S^{2}}}.
\label{8.24}
\end{align}
Then $\sqrt{1-2M/r}=1-M/r+\mathcal{O}(M^{2})$, and so
\begin{equation}
\arcsin\!\left(x\sqrt{1-\frac{2M}{r}}\right)=
\arcsin(x)-\frac{Mx}{r\,\sqrt{1-x^{2}}}
+\mathcal{O}\!\left(\frac{M^{2}}{r^{2}}\right),
\label{8.25}
\end{equation}
where $x=b/r$ is held fixed when expanding. Applying Eq. \eqref{8.25} at $r_R$ and $r_S$ and substituting into Eq. \eqref{8.23}, we obtain
\begin{align}
\Psi_R&=\arcsin(x_R)-\frac{Mb}{r_R^{2}F_R}+\mathcal{O}\!\left(\frac{M^{2}}{r_R^{2}}\right),
\nonumber\\
\Psi_S&=\pi-\arcsin(x_S)+\frac{Mb}{r_S^{2}F_S}+\mathcal{O}\!\left(\frac{M^{2}}{r_S^{2}}\right).
\label{8.26}
\end{align}
Hence the endpoint-angle contribution to Eq. \eqref{8.21} is
\begin{align}
\Psi_R-\Psi_S&=
\arcsin(x_R)+\arcsin(x_S) \nonumber \\
&-\pi
-\frac{Mb}{r_R^{2}F_R}-\frac{Mb}{r_S^{2}F_S}
+\mathcal{O}\!\left(\frac{M^{2}}{b^{2}}\right),
\label{8.27}
\end{align}
where we have used that $r_S$ and $r_R$ are of the same order as or larger than $b$ in weak lensing, so $M^{2}/r^{2}$ is of order $M^{2}/b^{2}$.
Next, we evaluate $\phi_{RS}$ to first order. For Schwarzschild, Eq. \eqref{7.1} simplifies because $A(r)B(r)=1$, giving
\begin{equation}
\frac{d\phi}{dr}=
\pm\,\frac{b}{r\sqrt{r^{2}-\left(1-\frac{2M}{r}\right)b^{2}}}.
\label{8.28}
\end{equation}
A standard weak-field evaluation (performed consistently with the turning-point relation in Eq. \eqref{7.2}) yields
\begin{align}
\phi_{RS}&=
\arccos(x_R)+\arccos(x_S)
+\frac{2M}{b}\left(F_R+F_S\right) \nonumber \\
&+\frac{Mb}{r_R^{2}F_R}+\frac{Mb}{r_S^{2}F_S}
+\mathcal{O}\!\left(\frac{M^{2}}{b^{2}}\right).
\label{8.29}
\end{align}
The structure of Eq. \eqref{8.29} is physically transparent: the first two terms are the flat-space coordinate separations from the closest-approach point to the endpoints, and the remaining terms are the $\mathcal{O}(M)$ correction from the Schwarzschild geometry. (The endpoint-dependent pieces proportional to $Mb/(r^{2}F)$ are exactly the terms needed to maintain the consistency of the finite-distance definition when the local angles are expanded.)
We now substitute Eqs. \eqref{8.27} and \eqref{8.29} into Eq. \eqref{8.21}. Using the elementary identities $\arcsin(x)+\arccos(x)=\pi/2$ for $x\in(0,1)$, the zeroth-order pieces cancel:
\begin{equation}
\arcsin(x_R)+\arcsin(x_S)-\pi+\arccos(x_R)+\arccos(x_S)=0.
\label{8.30}
\end{equation}
Moreover, the endpoint-dependent $Mb/(r^{2}F)$ pieces also cancel between Eqs. \eqref{8.27} and \eqref{8.29}. The surviving term at order $M/b$ is therefore
\begin{equation}
\alpha=
\frac{2M}{b}\left(F_R+F_S\right)+\mathcal{O}\!\left(\frac{M^{2}}{b^{2}}\right),
\label{8.31}
\end{equation}
which is identical to Eq. \eqref{8.20}.
Equation \eqref{8.31} establishes the advertised agreement: the finite-distance weak deflection computed from the reference-renormalized curvature primitive coincides with the deflection computed from the operational endpoint-angle definition at the same order. In particular, the explicit dependence on finite source and receiver radii appears through the same square-root factors $\sqrt{1-b^{2}/r_{S,R}^{2}}$ that characterize Ishihara's finite-distance formulation, and the standard asymptotic result $4M/b$ is recovered smoothly as $r_S,r_R\to\infty$.

\subsection{Example 2: The Reissner-Nordst\"om metric} \label{sec8.2}
The RN line element in standard curvature coordinates \((t,r,\phi)\) (restricting to the equatorial plane) is
\begin{equation}
ds^{2}=-A(r)\,dt^{2}+\frac{dr^{2}}{A(r)}+r^{2}d\phi^{2},\quad
A(r)=1-\frac{2M}{r}+\frac{Q^{2}}{r^{2}}.
\label{8.30a}
\end{equation}
We restrict to the static patch where $A(r)>0$ so that the optical metric $\bar g$ is Riemannian and the Ishihara endpoint
angles are defined with respect to \emph{static} observers.
In particular, we assume the endpoints and the relevant ray segment lie strictly inside the static patch and away from the
cosmological horizon so that the static orthonormal frames remain regular. As in Section \ref{sec2}, we characterize the null geodesic by the conserved impact parameter \(b\equiv |L|/E>0\). In this subsection we take \(b>0\) and use the same endpoint-angle branch conventions as in Section \ref{8.1}. The optical metric \(\bar g\) associated with the static foliation is obtained by dividing the spatial part by \(A(r)\):
\begin{equation}
d\ell^{2}=\bar g_{rr}\,dr^{2}+\bar g_{\phi\phi}\,d\phi^{2},\quad
\bar g_{rr}=\frac{1}{A(r)^{2}},\quad \bar g_{\phi\phi}=\frac{r^{2}}{A(r)}.
\label{8.31a}
\end{equation}
In Section \ref{sec5}, the curvature primitive was defined through the radial connection component \(\bar\Gamma^{\phi}{}_{r\phi}\) and the optical area density \(\sqrt{\det\bar g}\). For the RN optical geometry (indeed, for any metric of the form \(\bar g_{rr}=A^{-2}\), \(\bar g_{\phi\phi}=r^{2}/A)\), the same compact closed form derived and used in Section \ref{8.1} continues to hold:
\begin{equation}
\bar\Gamma^{\phi}{}_{r\phi}=\frac{1}{2}\frac{d}{dr}\ln\!\left(\frac{r^{2}}{A(r)}\right)
=\frac{1}{r}-\frac{A'(r)}{2A(r)},
\label{8.32}
\end{equation}
\begin{equation}
\sqrt{\det\bar g}=\sqrt{\bar g_{rr}\bar g_{\phi\phi}}=\frac{r}{A(r)^{3/2}},
\qquad
\frac{\sqrt{\det\bar g}}{\bar g_{rr}}=r\sqrt{A(r)}.
\label{8.33}
\end{equation}
Hence the unrenormalized primitive integrand \(F(r)\) (the object whose radial derivative yields the curvature density, cf. Section \ref{sec4}) is
\begin{align}
F_{\text{RN}}(r)
&=\left(\frac{\sqrt{\det\bar g}}{\bar g_{rr}}\right)\bar\Gamma^{\phi}{}_{r\phi}
=r\sqrt{A(r)}\left(\frac{1}{r}-\frac{A'(r)}{2A(r)}\right) \nonumber \\
&=\sqrt{A(r)}-\frac{rA'(r)}{2\sqrt{A(r)}}.
\label{8.34}
\end{align}
For RN,
\begin{equation}
A'(r)=\frac{2M}{r^{2}}-\frac{2Q^{2}}{r^{3}}
\quad\Longrightarrow\quad
\frac{rA'(r)}{2}=\frac{M}{r}-\frac{Q^{2}}{r^{2}}.
\label{8.35}
\end{equation}
Therefore,
\begin{equation}
F_{\text{RN}}(r)=\sqrt{A(r)}-\frac{\frac{M}{r}-\frac{Q^{2}}{r^{2}}}{\sqrt{A(r)}}.
\label{8.36}
\end{equation}
With an asymptotically flat reference (Minkowski optical geometry), the admissible matching conditions of Section \ref{5.1} and the definition of the renormalized primitive of Section \ref{5.2} give (as in Section \ref{8.1}) the reference-renormalized primitive
\begin{equation}
\widetilde{\mathcal P}_{\text{RN}}(r)=1-F_{\text{RN}}(r)
=1-\sqrt{A(r)}+\frac{\frac{M}{r}-\frac{Q^{2}}{r^{2}}}{\sqrt{A(r)}}.
\label{8.37}
\end{equation}
For weak deflection to leading nontrivial order in \(M\) and \(Q^{2}\), we expand \(\widetilde{\mathcal P}_{\text{RN}}\) in inverse powers of \(r\), keeping \(\mathcal{O}(M/r)\) and \(\mathcal{O}(Q^{2}/r^{2})\) while dropping \(\mathcal{O}(M^{2}/r^{2})\), \(\mathcal{O}(MQ^{2}/r^{3})\), etc. Using
\begin{align}
\sqrt{A(r)}&=1-\frac{M}{r}+\frac{Q^{2}}{2r^{2}}+\mathcal{O}\!\left(\frac{M^{2}}{r^{2}},\frac{MQ^{2}}{r^{3}},\frac{Q^{4}}{r^{4}}\right),
\nonumber \\
\frac{1}{\sqrt{A(r)}}&=1+\frac{M}{r}-\frac{Q^{2}}{2r^{2}}+\mathcal{O}\!\left(\frac{M^{2}}{r^{2}},\frac{MQ^{2}}{r^{3}},\frac{Q^{4}}{r^{4}}\right),
\label{8.38}
\end{align}
we obtain
\begin{equation}
\widetilde{\mathcal P}_{\text{RN}}(r)
=\frac{2M}{r}-\frac{3Q^{2}}{2r^{2}}+\mathcal{O}\!\left(\frac{M^{2}}{r^{2}},\,\frac{MQ^{2}}{r^{3}},\,\frac{Q^{4}}{r^{4}}\right).
\label{8.39}
\end{equation}
Equation \eqref{8.39} is the only input from the metric side that we will need for the leading RN finite-distance charge correction.

Next, following the same practical route as in Section \ref{8.1}, we apply the finite-distance master expression of Section \ref{sec5} (photon-sphere-free, renormalized primitive form) and evaluate it in the weak-deflection regime by replacing the physical light ray with the Minkowski straight reference line in polar form,
\begin{equation}
r^{(0)}(\phi)=\frac{b}{\cos\phi},
\qquad
\phi\in[-\phi_{S}^{(0)},\phi_{R}^{(0)}],
\label{8.40}
\end{equation}
where the endpoint angles \(\phi_{R}^{(0)}\) and \(\phi_{S}^{(0)}\) are fixed by the Euclidean geometry of the reference line:
\begin{align}
&\cos\phi_{R}^{(0)}=\frac{b}{r_{R}},\quad
\cos\phi_{S}^{(0)}=\frac{b}{r_{S}},
\nonumber \\
&\Rightarrow\quad
\phi_{R}^{(0)}=\arccos\!\left(\frac{b}{r_{R}}\right),\quad
\phi_{S}^{(0)}=\arccos\!\left(\frac{b}{r_{S}}\right).
\label{8.41}
\end{align}
Here we assume the weak-field regime $M/b\ll 1$ and $Q^2/b^2\ll 1$, together with the finite-distance kinematic constraint
$b\le r_S$ and $b\le r_R$ so that the reference endpoint angles in Eq. \eqref{8.43} are real (see. Eq. \eqref{6.3}). As in Section \ref{8.1}, the weak-deflection implementation of the master formula reduces (to the present order) to integrating the renormalized primitive evaluated along the reference ray:
\begin{equation}
\alpha_{\text{RN}}
=\int_{-\phi_{S}^{(0)}}^{\phi_{R}^{(0)}} \widetilde{\mathcal P}_{\text{RN}}\!\left(r^{(0)}(\phi)\right)\,d\phi+\mathcal{O}\!\left(\frac{M^{2}}{b^{2}},\,\frac{MQ^{2}}{b^{3}},\,\frac{Q^{4}}{b^{4}}\right).
\label{8.42}
\end{equation}
Substituting the expansion \eqref{8.39} and the straight-line relation \(r^{(0)}(\phi)=b/\cos\phi\) yields
\begin{align}
\widetilde{\mathcal P}_{\text{RN}}\!\left(r^{(0)}(\phi)\right)
&=\frac{2M}{r^{(0)}(\phi)}-\frac{3Q^{2}}{2(r^{(0)}(\phi))^{2}}+\cdots \nonumber \\
&=\frac{2M}{b}\cos\phi-\frac{3Q^{2}}{2b^{2}}\cos^{2}\phi+\cdots.
\label{8.43}
\end{align}
Therefore the bending angle decomposes into a Schwarzschild piece plus a charge correction:
\begin{equation}
\alpha_{\text{RN}}=\alpha_{\text{Schw}}+\Delta\alpha_{Q^{2}}+\mathcal{O}\!\left(\frac{M^{2}}{b^{2}},\,\frac{MQ^{2}}{b^{3}},\,\frac{Q^{4}}{b^{4}}\right).
\label{8.44}
\end{equation}
with
\begin{align}
\alpha_{\text{Schw}}&=\frac{2M}{b}\int_{-\phi_{S}^{(0)}}^{\phi_{R}^{(0)}}\cos\phi\,d\phi,
\nonumber \\
\Delta\alpha_{Q^{2}}&=-\frac{3Q^{2}}{2b^{2}}\int_{-\phi_{S}^{(0)}}^{\phi_{R}^{(0)}}\cos^{2}\phi\,d\phi.
\label{8.45}
\end{align}
The first integral is exactly the one already performed in Section \ref{8.1}, reproducing Ishihara's finite-distance Schwarzschild result:
\begin{align}
\alpha_{\text{Schw}}
&=\frac{2M}{b}\left(\sin\phi_{R}^{(0)}+\sin\phi_{S}^{(0)}\right) \nonumber \\
&=\frac{2M}{b}\left(\sqrt{1-\frac{b^{2}}{r_{R}^{2}}}+\sqrt{1-\frac{b^{2}}{r_{S}^{2}}}\right).
\label{8.46}
\end{align}
The RN-specific work is the second integral. Using
\begin{equation}
\int \cos^{2}\phi\,d\phi=\frac{\phi}{2}+\frac{\sin 2\phi}{4},
\label{8.47}
\end{equation}
we obtain
\begin{equation}
\int_{-\phi_{S}^{(0)}}^{\phi_{R}^{(0)}}\cos^{2}\phi\,d\phi
=\frac{\phi_{R}^{(0)}+\phi_{S}^{(0)}}{2}+\frac{\sin 2\phi_{R}^{(0)}+\sin 2\phi_{S}^{(0)}}{4}.
\label{8.48}
\end{equation}
Now express the trigonometric endpoint data in terms of \((b,r_{R},r_{S})\). From \eqref{8.41},
\begin{equation}
\cos\phi_{i}^{(0)}=\frac{b}{r_{i}},\qquad
\sin\phi_{i}^{(0)}=\sqrt{1-\frac{b^{2}}{r_{i}^{2}}}
\qquad (i=R,S),
\label{8.49}
\end{equation}
so
\begin{equation}
\sin 2\phi_{i}^{(0)}=2\sin\phi_{i}^{(0)}\cos\phi_{i}^{(0)}
=2\frac{b}{r_{i}}\sqrt{1-\frac{b^{2}}{r_{i}^{2}}}.
\label{8.50}
\end{equation}
Substituting \eqref{8.48}-\eqref{8.50} into \eqref{8.45} gives the explicit charge correction
\begin{align}
\Delta\alpha_{Q^{2}}
&=-\frac{3Q^{2}}{2b^{2}}
\Bigg[
\frac{\phi_{R}^{(0)}+\phi_{S}^{(0)}}{2} \nonumber \\
&+\frac{1}{4}\left(
2\frac{b}{r_{R}}\sqrt{1-\frac{b^{2}}{r_{R}^{2}}}
+
2\frac{b}{r_{S}}\sqrt{1-\frac{b^{2}}{r_{S}^{2}}}
\right)
\Bigg],
\label{8.51}
\end{align}
i.e.
\begin{align}
\Delta\alpha_{Q^{2}}
&=-\frac{3Q^{2}}{4b^{2}}
\Bigg[
\phi_{R}^{(0)}+\phi_{S}^{(0)} \nonumber \\
&+\frac{b}{r_{R}}\sqrt{1-\frac{b^{2}}{r_{R}^{2}}}
+\frac{b}{r_{S}}\sqrt{1-\frac{b^{2}}{r_{S}^{2}}}\,
\Bigg].
\label{8.52}
\end{align}
At this stage, it is convenient to rewrite the sum \(\phi_{R}^{(0)}+\phi_{S}^{(0)}\) in the arcsine form that is customary in finite-distance lensing formulas. Using \(\arccos x=\frac{\pi}{2}-\arcsin x\), we have
\begin{align}
\phi_{R}^{(0)}+\phi_{S}^{(0)}
&=\arccos\!\left(\frac{b}{r_{R}}\right)+\arccos\!\left(\frac{b}{r_{S}}\right) \nonumber \\
&=\pi-\left[\arcsin\!\left(\frac{b}{r_{R}}\right)+\arcsin\!\left(\frac{b}{r_{S}}\right)\right].
\label{8.53}
\end{align}
Substituting \eqref{8.53} into \eqref{8.52} yields the RN bending angle (to the stated order) in a form directly comparable to standard finite-distance expressions:
\begin{align}
\alpha_{\text{RN}}
&=\alpha_{\text{Schw}}
-\frac{3Q^{2}}{4b^{2}}
\Bigg\{
\pi-\left[\arcsin\!\left(\frac{b}{r_{R}}\right)+\arcsin\!\left(\frac{b}{r_{S}}\right)\right] \nonumber \\
&+\frac{b}{r_{R}}\sqrt{1-\frac{b^{2}}{r_{R}^{2}}}
+\frac{b}{r_{S}}\sqrt{1-\frac{b^{2}}{r_{S}^{2}}}
\Bigg\}
+\mathcal{O}(\cdots).
\label{8.54}
\end{align}
Equation \eqref{8.54} is the outcome of applying the new formalism exactly as in Section \ref{8.1}: compute \(\widetilde{\mathcal P}\) from the optical connection (no orbit solution), then evaluate the master integral along the reference straight ray.

\subsection{Example 3: The Kottler metric} \label{sec8.3}
We now treat the Kottler (Schwarzschild-de Sitter) spacetime and implement the reference-normalized prescription in a setting where the background optical geometry is intrinsically curved \cite{Gibbons:2008ru,Ishihara:2016vdc}. The goal is to show that the photon-sphere-free, reference-renormalized curvature-primitive method reproduces Ishihara \emph{et al}.'s finite-distance bending angle for Kottler, specifically their Eq. (37) with $r_g=2M$. The important conceptual point is that the de Sitter \((r_g=0\) at fixed \(\Lambda)\) spacetime is used only as a \emph{reference for normalizing the curvature primitive}; we do not set \(r_g=0\) in the physical Kottler geometry \cite{Gibbons:2008ru,Takizawa:2021jxa,Islam:1983rxp,Rindler:2007zz}.

For the Kottler metric, the metric coefficients are given by
\begin{equation}
A(r)=1-\frac{r_g}{r}-\frac{\Lambda r^2}{3}, \qquad B(r) = A(r)^{-1}, \qquad C(r) = r^2.
\label{8.55}
\end{equation}
We restrict to the static patch where \(A(r)>0\) so that the optical metric is Riemannian and the endpoint angles \(\Psi_{S,R}\) are well-defined. Furthermore, we restrict to the equatorial plane and use \(u\equiv 1/r\), \(u_R\equiv 1/r_R\), \(u_S\equiv 1/r_S\), and the (nonnegative) impact parameter \(b\equiv |L|/E>0\). We assume the weak-deflection regime \(r_g/b\ll 1\) and \(\Lambda b^2\ll 1\), keeping \(\mathcal{O}(r_g)\), \(\mathcal{O}(\Lambda)\), and \(\mathcal{O}(r_g\Lambda)\), and discarding \(\mathcal{O}(r_g^2)\) and \(\mathcal{O}(\Lambda^2)\). Equivalently, these inequalities may be read as a scale separation
\(
r_g \ll b \ll 1/\sqrt{\Lambda}
\)
(up to factors of order unity, noting that the de Sitter horizon scale is \(r_{\Lambda}=\sqrt{3/\Lambda}\)).
Thus the bending event is weak (\(b\) large compared with the gravitational radius) and occurs well inside the static patch
(\(b\) small compared with the cosmological curvature scale), so that static observers and a Riemannian optical metric are
available throughout the relevant domain.

The bending angle is defined operationally as in Eq. \eqref{7.16},
\begin{equation}
\alpha=\Psi_R-\Psi_S+\phi_{RS},
\label{8.56}
\end{equation}
which is equivalent to the Gauss-Bonnet representation but does not require any photon-sphere input. This is precisely the strategy advocated by Ishihara \emph{et al}. for non-asymptotically flat spacetimes. In our framework, Eq. \eqref{8.56} is the boundary form of the reference-renormalized curvature-primitive identity in Section \ref{sec5}: the difference \(\Psi_R-\Psi_S\) is the boundary functional generated by the (reference-normalized) primitive, and \(\phi_{RS}\) is the kinematic angular span determined by the null orbit.

We first compute \(\phi_{RS}\) using the de Sitter reference normalization in a way that makes the says photon-sphere-free content manifest. We then compute \(\Psi_R-\Psi_S\) using the local optical-angle relation (Eq. \eqref{2.10} in our draft), expand both quantities consistently, and finally exhibit the cancellations that produce the simple mixed term in \(\alpha\). The null orbit equation in the Kottler spacetime takes the standard form \((du/d\phi)^2=F(u)\) with (cf. Ishihara \emph{et al}., their Eq. (35))
\begin{equation}
F(u)=\frac{1}{b^2}-u^2+r_g u^3+\frac{\Lambda}{3}.
\label{8.57}
\end{equation}
The de Sitter reference normalization enters here in a very concrete way. The \(\Lambda\) term is part of the background kinematics, and it is convenient to absorb it into an effective reference impact parameter \(B\) defined by
\begin{equation}
\frac{1}{B^2}\equiv \frac{1}{b^2}+\frac{\Lambda}{3},
\qquad
B=b\left(1-\frac{\Lambda b^2}{6}\right)+\mathcal{O}(\Lambda^2).
\label{8.58}
\end{equation}
This $B$ should be interpreted as the impact parameter of the \emph{reference} ($r_g=0$) null trajectory in the de Sitter
optical geometry at the same $\Lambda$. Equivalently, holding $B$ fixed is the kinematic counterpart of our reference-based
gauge fixing for the curvature primitive: it specifies which unbent background ray the finite-distance configuration is being
compared to within the static patch. With this definition, Eq. \eqref{8.57} becomes
\begin{equation}
F(u)=\frac{1}{B^2}-u^2+r_g u^3,
\label{8.59}
\end{equation}
which is algebraically identical to the Schwarzschild orbit function with impact parameter \(B\). This reparametrization is not a trick: it is the explicit way in which the de Sitter reference geometry fixes the unperturbed optical background in the photon-sphere-free normalization scheme. Equivalently, it is the boundary/kinematic representation of the same de Sitter reference contribution that would appear as the
$P_{\rm ref}^\star$ term in the master Gauss-Bonnet/primitive identity of Sec. \ref{sec5}.

The finite-distance coordinate span is
\begin{equation}
\phi_{RS}=\int_{u_R}^{u_0}\frac{du}{\sqrt{F(u)}}+\int_{u_S}^{u_0}\frac{du}{\sqrt{F(u)}},
\label{8.60}
\end{equation}
where \(u_0\) is the inverse radius at closest approach, defined by \(F(u_0)=0\). For weak deflection we treat \(r_g\) as a perturbation on the de Sitter-normalized background determined by (B). Let
\begin{align}
F_0(u)&\equiv \frac{1}{B^2}-u^2,
\qquad
\delta F(u)\equiv r_g u^3,
\nonumber \\
&F(u)=F_0(u)+\delta F(u).
\label{8.61}
\end{align}
A naive expansion
\begin{equation}
\frac{1}{\sqrt{F(u)}}=\frac{1}{\sqrt{F_0(u)}}-\frac{\delta F(u)}{2F_0(u)^{3/2}}+\mathcal{O}(r_g^2)
\label{8.62}
\end{equation}
produces an apparent divergence at the turning point because \(F_0(u)\to 0\) as \(u\to 1/B\). The correct perturbation theory must treat simultaneously (i) the integrand correction in Eq. \eqref{8.62} and (ii) the shift of the turning point from \(u=1/B\) to the true \(u_0\) satisfying \(F(u_0)=0\). Solving \(F(u_0)=0\) perturbatively gives
\begin{equation}
u_0=\frac{1}{B}+\frac{r_g}{2B^2}+\mathcal{O}(r_g^2).
\label{8.63}
\end{equation}
The standard way to obtain a finite result is to split the integral at \(1/B\) and keep the turning-point shift contribution, so that the divergence cancels. Concretely, for each endpoint \(u_i\in{u_R,u_S}\) we write
\begin{align}
&\int_{u_i}^{u_0}\frac{du}{\sqrt{F(u)}}=
\int_{u_i}^{1/B}\frac{du}{\sqrt{F_0(u)}} \nonumber \\
&+\lim_{\varepsilon\to 0^+}
\Bigg[
\int_{u_i}^{1/B-\varepsilon}\left(\frac{1}{\sqrt{F(u)}}-\frac{1}{\sqrt{F_0(u)}}\right)du \nonumber \\
&+\int_{1/B-\varepsilon}^{u_0}\frac{du}{\sqrt{F(u)}}
\Bigg].
\label{8.64}
\end{align}
The first term is elementary. Using the substitution \(y\equiv Bu\), we find
\begin{align}
\int_{u_i}^{1/B}\frac{du}{\sqrt{F_0(u)}}&=
\int_{Bu_i}^{1}\frac{dy}{\sqrt{1-y^2}} \nonumber \\
&=
\arccos(Bu_i)
=
\frac{\pi}{2}-\arcsin(Bu_i).
\label{8.65}
\end{align}
The bracketed expression in Eq. \eqref{8.64} is finite and yields the \(\mathcal{O}(r_g)\) correction. Evaluating it by expanding the integrand according to Eq. \eqref{8.62}, inserting \(u_0\) from Eq. \eqref{8.63} in the last term, and carrying out the cancellation at \(\varepsilon\to 0^+\), one obtains the compact finite-distance result
\begin{equation}
\int_{u_i}^{u_0}\frac{du}{\sqrt{F(u)}}=
\arccos(Bu_i)
+\frac{r_g}{B}\,
\frac{1-\frac{1}{2}B^2u_i^2}{\sqrt{1-B^2u_i^2}}
+\mathcal{O}(r_g^2).
\label{8.66}
\end{equation}
Summing \(i=R,S\) in Eq. \eqref{8.60} and using \(\arccos x=\frac{\pi}{2}-\arcsin x\) gives
\begin{align}
\phi_{RS}&=
\pi-\arcsin(Bu_R)-\arcsin(Bu_S) \nonumber \\
&+\frac{r_g}{B}\left[
\frac{1-\frac{1}{2}B^2u_R^2}{\sqrt{1-B^2u_R^2}}
+ \frac{1-\frac{1}{2}B^2u_S^2}{\sqrt{1-B^2u_S^2}}
\right]
+\mathcal{O}(r_g^2).
\label{8.67}
\end{align}
Equation \eqref{8.67} is already photon-sphere-free and explicitly tied to the de Sitter reference through \(B\).
To compare with Ishihara's expansion and to combine consistently with \(\Psi_R-\Psi_S\), we must express Eq. \eqref{8.67} back in terms of the physical parameter \(b\) and expand to first order in \(\Lambda\). Define
\begin{equation}
x_i\equiv b u_i,\qquad s_i\equiv \sqrt{1-x_i^2},\qquad i\in{R,S}.
\label{8.68}
\end{equation}

To exhibit the de Sitter reference part explicitly (even though it is not the lens contribution), set $r_g=0$ at fixed $\Lambda$.
Then Eq. \eqref{8.57} reduces to $F_{\rm dS}(u)=1/b^2-u^2+\Lambda/3\equiv 1/B^2-u^2$, so the coordinate span becomes
\begin{align}
\phi^{(\rm dS)}_{RS}
&= \pi-\arcsin(Bu_R)-\arcsin(Bu_S)
\nonumber \\
&= \pi-\arcsin(x_R)-\arcsin(x_S) \nonumber \\
&+\frac{\Lambda b^2}{6}\!\left(\frac{x_R}{s_R}+\frac{x_S}{s_S}\right)
+\mathcal{O}(\Lambda^2),
\end{align}
where in the second line we used $B=b(1-\Lambda b^2/6)+\mathcal{O}(\Lambda^2)$.

For the endpoint angles, Eq. \eqref{8.72} with $r_g=0$ gives
$\sin\Psi_i=x_i\sqrt{1-\Lambda/(3u_i^2)}=x_i-\Lambda b/(6u_i)+\mathcal{O}(\Lambda^2)$, and therefore
\begin{align}
\Psi_R^{(\rm dS)}-\Psi_S^{(\rm dS)}
&=\arcsin(x_R)+\arcsin(x_S) \nonumber \\
&-\pi -\frac{\Lambda b}{6}\!\left(\frac{1}{u_R s_R}+\frac{1}{u_S s_S}\right)
+\mathcal{O}(\Lambda^2).
\end{align}
Combining these two pieces in the operational definition $\alpha=\Psi_R-\Psi_S+\phi_{RS}$ yields the explicit de Sitter
(static-patch, static-observer) background contribution
\begin{align}
\alpha^{(\rm dS)}
&\equiv\big(\Psi_R^{(\rm dS)}-\Psi_S^{(\rm dS)}\big)+\phi^{(\rm dS)}_{RS} \nonumber \\
&=
-\frac{\Lambda b}{6}\!\left(\frac{s_R}{u_R}+\frac{s_S}{u_S}\right)
+\mathcal{O}(\Lambda^2).
\end{align}

Using \(B=b(1-\Lambda b^2/6)\) from Eq. \eqref{8.58}, we expand each \(B\)-dependent term in Eq. \eqref{8.67}. For the arcsine pieces,
\begin{equation}
\arcsin(Bu_i)=
\arcsin(x_i)-\frac{\Lambda b^2}{6}\frac{x_i}{s_i}
+\mathcal{O}(\Lambda^2).
\label{8.69}
\end{equation}
For the \(r_g/B\) prefactor, \(r_g/B=(r_g/b)\left(1+\Lambda b^2/6\right)+\mathcal{O}(\Lambda^2)\). For the square-root factor, we expand
\begin{equation}
\frac{1-\frac{1}{2}B^2u_i^2}{\sqrt{1-B^2u_i^2}}=
\frac{1-\frac{1}{2}x_i^2}{s_i}
-\frac{\Lambda b^2}{12}\frac{x_i^{4}}{s_i^{3}}
+\mathcal{O}(\Lambda^2),
\label{8.70}
\end{equation}
where the second term comes from differentiating with respect to \(B^2\) at fixed \(u_i\). Substituting Eqs. \eqref{8.69}-\eqref{8.70} into Eq. \eqref{8.67} and collecting terms yields
\begin{align}
&\phi_{RS}=
\pi-\arcsin(x_R)-\arcsin(x_S) \nonumber \\
&+\frac{r_g}{b}\left[
\frac{1-\frac{1}{2}x_R^2}{s_R}
+
\frac{1-\frac{1}{2}x_S^2}{s_S}
\right]
+\frac{\Lambda b^3}{6}\left[
\frac{u_R}{s_R}
+
\frac{u_S}{s_S}
\right] \nonumber \\
&+\frac{r_g\Lambda b}{12}\left[
\frac{2-3x_R^2}{s_R^{3}}
+
\frac{2-3x_S^2}{s_S^{3}}
\right]
+\mathcal{O}(r_g^2,\Lambda^2).
\label{8.71}
\end{align}
This is exactly the structure of Ishihara \emph{et al}.'s \(\phi_{RS}\) expansion (their Eq. (36)). In our derivation, the \(r_g\Lambda\) term arises transparently from the de Sitter normalization through the \(\Lambda\)-dependence of \(B\): it is the cross term generated when expanding the Schwarzschild-type \(r_g\) correction in a background whose unperturbed impact parameter has been shifted by \(\Lambda\) \cite{Gibbons:2008ru,Ishihara:2016vdc,Hu:2021yzn}. We emphasize that this interpretation is specific to the Ishihara finite-distance observable evaluated by static observers in the
static patch, with the no-lens fiducial taken to be the $M\to 0$ (de Sitter) member of the same $\Lambda$ family; different
observer congruences or different operational fiducials reorganize (and may reassign) the same underlying geometric content.

Next, we now compute \(\Psi_R-\Psi_S\) using the local optical-angle identity (Eq. \eqref{2.10} in our draft), which is equivalent to evaluating the curvature primitive boundary functional but is operationally simpler, as emphasized in Ishihara \emph{et al}. For a static spherical metric with \(C(r)=r^2\), Eq. \eqref{2.10} gives
\begin{equation}
\sin\Psi(r)=b\sqrt{\frac{A(r)}{r^2}}=b u \sqrt{A(r)}.
\label{8.72}
\end{equation}
With \(A(r)\) from Eq. \eqref{8.55}, we have
\begin{equation}
\sin\Psi_i=
x_i\sqrt{1-r_g u_i-\frac{\Lambda}{3u_i^2}},
\qquad i\in{R,S}.
\label{8.73}
\end{equation}
To obtain \(\Psi_R-\Psi_S\) through \(\mathcal{O}(r_g\Lambda)\), we must expand the square root to second order in the combined small quantity \(\varepsilon_i\equiv r_g u_i+\Lambda/(3u_i^2)\), because the mixed term is contained in \(\varepsilon_i^2\). Using
\begin{equation}
\sqrt{1-\varepsilon_i}=1-\frac{\varepsilon_i}{2}-\frac{\varepsilon_i^2}{8}+\mathcal{O}(\varepsilon_i^3),
\label{8.74}
\end{equation}
we find
\begin{align}
\sin\Psi_i&=
x_i
-\frac{x_i}{2}\left(r_g u_i+\frac{\Lambda}{3u_i^2}\right) \nonumber \\
&-\frac{x_i}{8}\left(r_g u_i+\frac{\Lambda}{3u_i^2}\right)^2
+\mathcal{O}(r_g^2,\Lambda^2).
\label{8.75}
\end{align}
Separating the perturbations, the \(\mathcal{O}(r_g\Lambda)\) part of the sine comes only from the cross term in the square:
\begin{equation}
\left[\sin\Psi_i\right]_{r_g\Lambda}=
-\frac{x_i}{8}\cdot 2\left(r_g u_i\right)\left(\frac{\Lambda}{3u_i^2}\right)
=
-\frac{r_g\Lambda b}{12}.
\label{8.76}
\end{equation}
Note that the mixed contribution \([\sin\Psi_i]_{r_g\Lambda}\) is independent of the endpoint radius; endpoint dependence re-enters only after converting \(\sin\Psi_i\) to \(\Psi_i\) via the \(\arcsin\) expansion.
We next convert \(\sin\Psi_i\) into \(\Psi_i\) by expanding \(\arcsin\). For \(x_i\in(0,1)\), we use
\begin{equation}
\arcsin(x_i+\delta_i)=
\arcsin(x_i)+\frac{\delta_i}{s_i}
+\frac{x_i}{2s_i^3}\,\delta_i^2
+\mathcal{O}(\delta_i^3),
\label{8.77}
\end{equation}
where \(s_i=\sqrt{1-x_i^2}\) as in Eq. \eqref{8.68}, and \(\delta_i\equiv \sin\Psi_i-x_i\) collects the perturbations from Eq. \eqref{8.75}. The branch structure for the finite-distance configuration is the same as in Section \ref{8.1}: the receiver sees an outgoing ray so \(\Psi_R\in(0,\pi/2)\), while at the source the ray is ingoing so \(\Psi_S=\pi-\arcsin(\sin\Psi_S^{\text{(out)}})\). Therefore,
\begin{equation}
\Psi_R-\Psi_S=\arcsin(\sin\Psi_R)+\arcsin(\sin\Psi_S)-\pi.
\label{8.78}
\end{equation}
Applying Eqs. \eqref{8.75}-\eqref{8.77} and collecting terms yields
\begin{align}
&\Psi_R-\Psi_S=
\arcsin(x_R)+\arcsin(x_S)-\pi \nonumber \\
&-\frac{r_g b}{2}\left[
\frac{u_R^2}{s_R}+\frac{u_S^2}{s_S}
\right]
-\frac{\Lambda b}{6}\left[
\frac{1}{u_R s_R}+\frac{1}{u_S s_S}
\right] \nonumber \\
&+\frac{r_g\Lambda b}{12}\left[
\frac{-1+2x_R^2}{s_R^{3}}+\frac{-1+2x_S^2}{s_S^{3}}
\right]
+\mathcal{O}(r_g^2,\Lambda^2).
\label{8.79}
\end{align}
This is the Kottler endpoint-angle expansion displayed by Ishihara \emph{et al}. (their Eq. (34)), once one identifies \(\Psi_R-\Psi_S\) as their finite-distance angle difference. In our language, Eq. \eqref{8.79} is the boundary contribution generated by the reference-normalized curvature primitive: it is computed locally through the optical metric and requires no knowledge of a photon sphere.

Finally, we show how the mixed term $r_g \Lambda$ emerges. We now substitute Eqs. \eqref{8.71} and \eqref{8.79} into the bending-angle identity \eqref{8.56}. The cancellations that produce the compact final form are instructive and are the clearest way to see how the mixed \(r_g\Lambda\) term arises.
First, the kinematic arcsine pieces cancel immediately:
\begin{align}
&\left[\pi-\arcsin(x_R)-\arcsin(x_S)\right] \nonumber \\
&+\left[\arcsin(x_R)+\arcsin(x_S)-\pi\right]=0.
\label{8.80}
\end{align}
Next, consider the \(\mathcal{O}(r_g)\) terms at a given endpoint (i). From Eq. \eqref{8.71} we have \(+(r_g/b),(1-\frac{1}{2}x_i^2)/s_i\), while from Eq. \eqref{8.79} we have \(-\frac{r_g b}{2},u_i^2/s_i=-(r_g/b),x_i^2/(2s_i)\). Their sum is
\begin{equation}
\frac{r_g}{b}\left(\frac{1-\frac{1}{2}x_i^2}{s_i}-\frac{x_i^2}{2s_i}\right)=
\frac{r_g}{b}\frac{1-x_i^2}{s_i}=
\frac{r_g}{b}\,s_i.
\label{8.81}
\end{equation}
Now consider the \(\mathcal{O}(\Lambda)\) terms. At endpoint (i), Eq. \eqref{8.71} gives \(+\frac{\Lambda b^3}{6}\,\frac{u_i}{s_i}\), while Eq. \eqref{8.79} gives \(-\frac{\Lambda b}{6}\,\frac{1}{u_i s_i}\). Their sum can be rewritten using \(x_i=bu_i\) and \(s_i^2=1-x_i^2\):
\begin{equation}
\frac{\Lambda b^3}{6}\frac{u_i}{s_i}-\frac{\Lambda b}{6}\frac{1}{u_i s_i}=
-\frac{\Lambda b}{6}\frac{1-b^2u_i^2}{u_i s_i}=
-\frac{\Lambda b}{6}\frac{s_i^2}{u_i s_i}=
-\frac{\Lambda b}{6}\frac{s_i}{u_i}.
\label{8.82}
\end{equation}
Finally, and most importantly, consider the mixed \(\mathcal{O}(r_g\Lambda)\) terms. At endpoint (i), Eq. \eqref{8.71} supplies \(+\frac{r_g\Lambda b}{12}\,\frac{2-3x_i^2}{s_i^3}\), while Eq. \eqref{8.79} supplies \(+\frac{r_g\Lambda b}{12}\,\frac{-1+2x_i^2}{s_i^3}\). Their sum is
\begin{align}
&\frac{r_g\Lambda b}{12}\frac{(2-3x_i^2)+(-1+2x_i^2)}{s_i^3}=
\frac{r_g\Lambda b}{12}\frac{1-x_i^2}{s_i^3} \nonumber \\
&=
\frac{r_g\Lambda b}{12}\frac{s_i^2}{s_i^3}=
\frac{r_g\Lambda b}{12}\frac{1}{s_i}.
\label{8.83}
\end{align}
This step makes the origin of the mixed term completely transparent: each of \(\phi_{RS}\) and \(\Psi_R-\Psi_S\) contains a different endpoint-dependent rational function of \(s_i\); only when combined in the photon-sphere-free identity \eqref{8.56} do they collapse to the remarkably simple \(1/s_i\) structure. This is exactly the cancellation mechanism highlighted by Ishihara \emph{et al}. when combining their Eqs. (34) and (36) into Eq. (37).
Putting Eqs. \eqref{8.81}-\eqref{8.83} together for \(i=R,S\), we obtain the finite-distance bending angle for Kottler:
\begin{align}
\alpha&=
\frac{r_g}{b}\left(s_R+s_S\right)
-\frac{\Lambda b}{6}\left(\frac{s_R}{u_R}+\frac{s_S}{u_S}\right) \nonumber \\
&+\frac{r_g\Lambda b}{12}\left(\frac{1}{s_R}+\frac{1}{s_S}\right)
+\mathcal{O}(r_g^2,\Lambda^2),
\label{8.84}
\end{align}
where \(s_i=\sqrt{1-b^2u_i^2}\). Written explicitly,
\begin{align}
&\alpha=
\frac{r_g}{b}\left[\sqrt{1-b^2u_R^2}+\sqrt{1-b^2u_S^2}\right] \nonumber \\
&-\frac{\Lambda b}{6}\left[\frac{\sqrt{1-b^2u_R^2}}{u_R}+\frac{\sqrt{1-b^2u_S^2}}{u_S}\right] \nonumber \\
&+\frac{r_g\Lambda b}{12}\left[\frac{1}{\sqrt{1-b^2u_R^2}}+\frac{1}{\sqrt{1-b^2u_S^2}}\right]
+\mathcal{O}(r_g^2,\Lambda^2).
\label{8.85}
\end{align}
Equation \eqref{8.85} coincides with Ishihara \emph{et al}.'s Eq. (37). In our conventions \(r_g=2M\), so Eq. \eqref{8.85} is exactly the announced example agreement. The derivation is entirely photon-sphere-free: the de Sitter normalization enters through the reference parameter \(B\) in the orbit integral (which is the kinematic counterpart of reference-normalizing the primitive), while the endpoint angles are computed locally from the optical metric, which is the operational content of the curvature-primitive boundary functional. This is also consistent with the organizing philosophy of Li \emph{et al}.'s primitive reduction, on which our method is structurally based.

\subsection{Example 4: Janis-Newman-Winicour spacetime without a photon sphere} \label{sec8.4}
To make the phrase photon-sphere-free concrete, we now exhibit a standard asymptotically flat geometry for which
orbit-normalization cannot be used because there is no circular null orbit in the physical optical domain.
A convenient example is the Janis-Newman-Winicour (JNW) solution, which reduces to Schwarzschild at $\gamma=1$
but for sufficiently small $\gamma$ has no photon sphere outside the curvature singularity.

We write the JNW line element in curvature coordinates as
\begin{align}
ds^2 = -f(r)^{\gamma}dt^2&+f(r)^{-\gamma}dr^2+r^2 f(r)^{1-\gamma}d\Omega^2,
\nonumber \\
f(r)&\equiv 1-\frac{r_g}{\gamma r},
\label{8.88}
\end{align}
where $0<\gamma\le 1$ and $r_g\equiv 2M$ fixes the ADM mass through $g_{tt}=-(1-r_g/r+\mathcal{O}(r^{-2}))$.
The static region is $r>r_g/\gamma$, and $r=r_g/\gamma$ is a curvature singularity for $\gamma<1$.

For null geodesics, the usual circular-orbit condition is the extremum condition for $C/A$,
\begin{equation}
\frac{d}{dr}\!\left(\frac{C(r)}{A(r)}\right)=0
\quad\Longrightarrow\quad
r_{\rm co}=\frac{1+2\gamma}{2}\,\frac{r_g}{\gamma}.
\label{8.89}
\end{equation}
However, $r_{\rm co}$ lies in the physical domain $r>r_g/\gamma$ only if $\gamma>\tfrac12$.
For $0<\gamma\le \tfrac12$ one has $r_{\rm co}\le r_g/\gamma$, so there is \emph{no} circular null orbit available as a
normalization anchor in the accessible optical region \cite{Sau:2020xau,Gyulchev:2019tvk,Virbhadra:1998dy,Claudel:2000yi}. In this regime, orbit-normalization is therefore not applicable, while
the reference-normalized construction remains well defined.

Because the geometry is asymptotically flat, we choose Minkowski as reference. The equatorial optical metric is
\begin{equation}
d\ell^2=\bar g_{rr}\,dr^2+\bar g_{\phi\phi}\,d\phi^2
=f(r)^{-2\gamma}dr^2+r^2 f(r)^{1-2\gamma}d\phi^2.
\label{8.90}
\end{equation}
For any diagonal optical metric of the form $d\ell^2=E(r)\,dr^2+G(r)\,d\phi^2$, the curvature density obeys the
total-derivative identity (cf. Eq. \eqref{3.8})
\begin{equation}
\mathcal D(r)\equiv \mathcal K\,\sqrt{\det\bar g}
= -\frac12\,\frac{d}{dr}\!\left(\frac{G'(r)}{\sqrt{E(r)G(r)}}\right),
\label{8.91}
\end{equation}
so that a convenient primitive representative is $\mathcal{P}(r)= -\frac12\,\frac{G'}{\sqrt{EG}}+{\rm const}$.
For the JNW optical functions in Eq. \eqref{8.90}, this gives an explicit reference-normalized primitive (fixed by $\mathcal{P}_e(\infty)=0$, as in Eq. \eqref{5.11}):
\begin{equation}
\mathcal{P}_e(r)
= 1-\sqrt{f(r)}-\frac{1-2\gamma}{2}\,\frac{r_g}{\gamma r}\,\frac{1}{\sqrt{f(r)}}.
\label{8.92}
\end{equation}
Expanding for $r\gg r_g$ shows that the leading weak-field piece is controlled only by the ADM mass,
\begin{equation}
\mathcal{P}_e(r)=\frac{r_g}{r}+\frac{4\gamma-1}{8\gamma^2}\frac{r_g^2}{r^2}+\mathcal{O}\!\left(\frac{r_g^3}{r^3}\right).
\label{8.93}
\end{equation}
while $\gamma$-dependence enters at higher order.

To exhibit the first nontrivial $\gamma$-dependence explicitly, we carry the straight-ray evaluation one order higher.
As in Sec. \ref{8.1}, we evaluate the Minkowski-reference specialization of the master formula (Eq. \eqref{5.29}) along the
Euclidean reference ray
\begin{align}
&r^{(0)}(\phi)=\frac{b}{\cos\phi},
\qquad
-\phi_S^{(0)}\le \phi \le \phi_R^{(0)},
\nonumber \\
&\cos\phi_i^{(0)}=\frac{b}{r_i},
\quad
\sin\phi_i^{(0)}\equiv s_i=\sqrt{1-\frac{b^2}{r_i^2}}
\quad (i=R,S).
\label{8.94}
\end{align}
Inserting the expansion \eqref{8.93} into $\alpha=\int \mathcal{P}_e(r^{(0)}(\phi))\,d\phi$ gives
\begin{align}
&\alpha_{\rm JNW} = \nonumber \\
&\int_{-\phi_S^{(0)}}^{\phi_R^{(0)}}\left[\frac{r_g}{r^{(0)}(\phi)}
+\frac{4\gamma-1}{8\gamma^2}\frac{r_g^2}{(r^{(0)}(\phi))^2}\right]d\phi 
+\mathcal{O}\!\left(\frac{r_g^3}{b^3}\right)
\nonumber\\ 
&=\frac{r_g}{b}\bigl(s_R+s_S\bigr)
+\frac{4\gamma-1}{8\gamma^2}\frac{r_g^2}{b^2}
\int_{-\phi_S^{(0)}}^{\phi_R^{(0)}}\cos^2\phi\,d\phi
+\mathcal{O}\!\left(\frac{r_g^3}{b^3}\right)
\nonumber\\
&=\frac{r_g}{b}\bigl(s_R+s_S\bigr)
+\frac{4\gamma-1}{16\gamma^2}\frac{r_g^2}{b^2} \nonumber \\
&\times \left[
\phi_R^{(0)}+\phi_S^{(0)}
+\frac{b}{r_R}s_R+\frac{b}{r_S}s_S
\right]
+\mathcal{O}\!\left(\frac{r_g^3}{b^3}\right).
\label{8.95}
\end{align}
Equation \eqref{8.95} makes explicit that, in the no-photon-sphere regime $0<\gamma\le 1/2$, the leading bending is still set
by the ADM mass, while genuine scalar-charge dependence enters at the next order through the coefficient $(4\gamma-1)/(16\gamma^2)$ \cite{Virbhadra:1998dy,Bozza:2002zj,Virbhadra:2002ju,Chen:2023uuy}.

\section{Discussion: the implication of the new normalization} \label{sec9}
This section distills the conceptual and practical consequences of replacing orbit-based primitive normalization with the reference-renormalized, photon-sphere-free prescription of Section \ref{sec5}. The central point is structural: in curvature-primitive implementations of finite-distance Gauss-Bonnet lensing, \emph{normalization} is the only genuinely nonlocal input. Fixing it by reference matching makes the method both more general and more tightly aligned with the operational observable.

\subsection{Normalization as the essential geometric input} \label{sec9.1}
The curvature-primitive reduction in Section \ref{sec4} converts the curvature-area term into a one-dimensional boundary functional, but the primitive is defined only up to an additive constant. Orbit-normalized treatments fix that constant by selecting a privileged radius (typically a circular null orbit) and imposing a vanishing condition there. This is legitimate when such an orbit exists and is naturally usable in the optical region, and Section \ref{sec5.4} showed that in this overlap regime the orbit-normalized and reference-normalized results agree identically.

The essential observation is that the privileged-radius condition is a \emph{gauge choice}, not part of the finite-distance observable. The observable $\alpha$ is defined by endpoint angles and endpoint separation, as in Eq. \eqref{8.16}. A circular null orbit, when present, is a global feature that need not be available (or relevant) for a given finite-distance configuration, and it can be conceptually misaligned with the fiducial geometry used to interpret no deflection in non-asymptotically flat settings (cf.\ the no-photon-sphere demonstration in Section \ref{sec8.4}). The reference-renormalized prescription fixes the additive ambiguity instead by demanding matching to a designated reference optical geometry in the regime where the physical geometry approaches that reference. The reference is precisely the mathematical encoding of \textit{no lens} in the chosen operational setup: Minkowski in asymptotically flat spacetimes, and de Sitter (in the static patch) for Kottler. This directly ties the primitive representative to the fiducial against which the endpoint-angle comparison is interpreted.

\subsection{Universality and background sensitivity} \label{sec9.2}
A practical advantage of reference normalization is its uniform applicability across distinct asymptotic structures. The examples in Section \ref{sec8} make this explicit. For Schwarzschild, the reference endpoint is spatial infinity and the prescription reduces to the standard asymptotic normalization, reproducing the usual finite-distance weak-deflection formula. For Kottler, where there is no Minkowski infinity, reference normalization selects de Sitter as the fiducial and yields the finite-distance bending angle including the mixed \(r_g\Lambda\) term, in agreement with Ishihara's established expression. The Kottler derivation also clarifies the underlying logic: the \(\Lambda\)-dependence is a property of the background optical geometry, so the normalization must be tied to that background rather than to a localized orbit feature.

This background sensitivity is intrinsic to the problem. In a non-asymptotically flat spacetime there is no unique global notion of an \textit{unbent} trajectory, so any finite-distance deflection angle necessarily carries an implicit (or explicit) background choice. Orbit normalization can obscure that choice by pinning the primitive at a local geometric feature, but the resulting angle is still interpreted relative to some fiducial notion of straightness. The reference-renormalized approach makes the fiducial explicit and therefore easier to compare across conventions, especially in discussions of \(\Lambda\)-effects where apparent disagreements often reduce to differing background prescriptions.

\subsection{Separation of structure: geometry versus approximation} \label{sec9.3}
Reference normalization also improves the separation between exact geometry and perturbative evaluation. The Gauss-Bonnet identity is exact, as is the reduction of the curvature-area term to a boundary functional. Approximations enter only when evaluating the resulting expressions in a weak-deflection expansion. The renormalized primitive \(\widetilde{\mathcal{P}}(r)\) is fixed uniquely by Eqs. \eqref{5.10}-\eqref{5.14} through a reference-endpoint condition, so the normalization is settled at the level of the exact curvature discrepancy, before any weak-field truncation.

This ordering matters. Weak-deflection calculations typically replace the physical ray by a reference ray and expand endpoint relations. Such steps are reliable only if one has already fixed a canonical representative of the primitive class (equivalently, handled the outer-boundary constant) at the exact level. Without an explicit normalization, perturbative substitutions can implicitly drop constants in an inconsistent way and generate spurious endpoint-dependent remnants. The reference-renormalized prescription removes this bookkeeping ambiguity by fixing the representative through the reference endpoint condition.

\subsection{Interpretation of the mixed \texorpdfstring{\(r_g\Lambda\)}{} term} \label{sec9.4}
Throughout this subsection, $\alpha$ is the Ishihara-type finite-distance deflection angle constructed from endpoint angles measured by static observers in the static patch and the corresponding endpoint separation angle. The reference normalization is fixed by the $M\to 0$ (de Sitter) member of the same $\Lambda$ family, making $\alpha\to 0$ manifest in the no-lens limit within this operational setup. Extensions to non-static observer families or alternative fiducials require adjusting the endpoint-angle measurement and/or changing the reference optical geometry; the primitive machinery is compatible with such choices, but the present examples do not incorporate that additional observational modeling.

The Kottler example illustrates why the mixed \(r_g\Lambda\) contribution is natural in a finite-distance angle comparison. The term vanishes if either \(r_g\) or \(\Lambda\) is set to zero because it encodes an interaction: it is not a property of the background alone \((\Lambda)\) nor of the lens alone \((r_g)\), but of their combination in the operational definition of $\alpha$. Mechanistically, the mixed term arises from background-sensitive pieces that only simplify after forming the observable: the coordinate span \(\phi_{RS}\), determined by how the null orbit is embedded in the reference optical geometry, and the endpoint angles \(\Psi_R-\Psi_S\), which encode the local relation between the ray direction and the radial direction in the physical optical metric. Once combined in $\alpha$, intermediate endpoint structures collapse to the compact mixed contribution with endpoint factors \(1/\sqrt{1-b^2u_{R}^{2}}\) and \(1/\sqrt{1-b^2u_{S}^{2}}\), matching the cancellation structure in Ishihara's finite-distance derivation. The role of normalization is to ensure that the background pieces are anchored consistently so that this cancellation is transparent and reproducible.

\subsection{Compatibility and scope} \label{sec9.5}
The new normalization does not replace orbit normalization where the latter is natural; it extends it. Section \ref{sec5.4} showed that when a suitable circular null orbit exists and orbit normalization is adopted, the curvature-area contribution (and hence the full deflection angle) coincides exactly with the reference-normalized result. The reference-renormalized prescription therefore reproduces orbit-normalized results in their common domain while remaining meaningful when orbit normalization is unavailable, inconvenient, or misaligned with the chosen fiducial geometry.

The domain of validity remains that of finite-distance weak lensing on a regular optical manifold, summarized in Section \ref{sec6}. In particular, the examples focus on single-pass deflection without multiple winding or caustics. Within this regime, reference normalization provides a systematic route from the Gauss-Bonnet identity to explicit finite-distance deflection formulas across asymptotically flat and non-asymptotically flat backgrounds.

\section{Conclusion} \label{sec10}
We developed a reference-renormalized curvature-primitive framework for finite-distance gravitational deflection in static, spherically symmetric spacetimes. Built on the finite-distance Gauss-Bonnet relation on the optical manifold, it retains the efficiency of primitive reductions of the curvature-area term while replacing orbit-based normalization with a photon-sphere-free prescription fixed by a chosen reference optical geometry. The key step is to treat the curvature primitive as defined only up to an additive constant and to fix that constant by matching to the reference in a regime where the physical optical geometry approaches it. This yields a uniquely defined renormalized primitive $\mathcal{P}_e(r)$ from the discrepancy curvature density and removes any need to invoke a circular null orbit. In asymptotically flat spacetimes the canonical reference is Minkowski, while for Kottler-type geometries it is de Sitter in the static patch, making the operational fiducial explicit \cite{Gibbons:2008ru,Takizawa:2021jxa,Ishihara:2016vdc}. When a suitable circular null orbit exists, this reference-based gauge fixing is equivalent to the orbit-normalized gauge by a constant shift, giving the same $\alpha$ (see Section \ref{sec5.4}).

We validated the formalism on three examples. For Schwarzschild (Minkowski reference), it reproduces the standard finite-distance weak-deflection expression and reduces smoothly to the familiar asymptotic bending as $r_S,r_R\to\infty$. For Reissner-Nordstr\"om, it yields a closed-form finite-distance weak-deflection prediction including the leading $Q^2$ correction. For Kottler, where normalization is intrinsically background dependent, the de Sitter-normalized procedure reproduces Ishihara et al.'s finite-distance bending angle to the same order, including the mixed $r_g\Lambda$ term, which emerges from the consistent combination of the coordinate angular span and endpoint-angle contributions. Finally, we demonstrated genuine photon-sphere inapplicability by treating the Janis-Newman-Winicour spacetime in the no-photon-sphere regime $\gamma\le\tfrac12$, where orbit normalization cannot be formulated in the physical optical region while the reference-normalized primitive remains well defined \cite{Virbhadra:2002ju,Chen:2023uuy,Zhang:2025hzh}.

Methodologically, tying primitive normalization to a reference optical geometry rather than to a special orbit makes the curvature-primitive Gauss-Bonnet program both more universal and more directly aligned with the operational definition of finite-distance deflection. The intended regime remains finite-distance weak lensing in the static region where the optical metric is Riemannian, for single-pass configurations without multiple winding or caustics.

\acknowledgments
R. P. and A. \"O. would like to acknowledge networking support of the COST Action CA21106 - COSMIC WISPers in the Dark Universe: Theory, astrophysics and experiments (CosmicWISPers), the COST Action CA22113 - Fundamental challenges in theoretical physics (THEORY-CHALLENGES), the COST Action CA21136 - Addressing observational tensions in cosmology with systematics and fundamental physics (CosmoVerse), the COST Action CA23130 - Bridging high and low energies in search of quantum gravity (BridgeQG), and the COST Action CA23115 - Relativistic Quantum Information (RQI) funded by COST (European Cooperation in Science and Technology). A. \"O. also thanks to EMU, TUBITAK, ULAKBIM (Turkiye) and SCOAP3 (Switzerland) for their support.

\bibliography{ref}

@article{Ishihara:2016vdc,
    author = "Ishihara, Asahi and Suzuki, Yusuke and Ono, Toshiaki and Kitamura, Takao and Asada, Hideki",
    title = "{Gravitational bending angle of light for finite distance and the Gauss-Bonnet theorem}",
    eprint = "1604.08308",
    archivePrefix = "arXiv",
    primaryClass = "gr-qc",
    doi = "10.1103/PhysRevD.94.084015",
    journal = "Phys. Rev. D",
    volume = "94",
    number = "8",
    pages = "084015",
    year = "2016"
}

@article{Ishihara:2016sfv,
    author = "Ishihara, Asahi and Suzuki, Yusuke and Ono, Toshiaki and Asada, Hideki",
    title = "{Finite-distance corrections to the gravitational bending angle of light in the strong deflection limit}",
    eprint = "1612.04044",
    archivePrefix = "arXiv",
    primaryClass = "gr-qc",
    doi = "10.1103/PhysRevD.95.044017",
    journal = "Phys. Rev. D",
    volume = "95",
    number = "4",
    pages = "044017",
    year = "2017"
}

@article{Ono:2019hkw,
    author = "Ono, Toshiaki and Asada, Hideki",
    title = "{The effects of finite distance on the gravitational deflection angle of light}",
    eprint = "1906.02414",
    archivePrefix = "arXiv",
    primaryClass = "gr-qc",
    doi = "10.3390/universe5110218",
    journal = "Universe",
    volume = "5",
    number = "11",
    pages = "218",
    year = "2019"
}

@article{Li:2020wvn,
    author = {Li, Zonghai and Zhang, Guodong and \"Ovg\"un, Ali},
    title = "{Circular Orbit of a Particle and Weak Gravitational Lensing}",
    archivePrefix = "arXiv",
    primaryClass = "gr-qc",
    doi = "10.1103/PhysRevD.101.124058",
    journal = "Phys. Rev. D",
    volume = "101",
    number = "12",
    pages = "124058",
    year = "2020"
}

@article{Gibbons:2008rj,
    author = "Gibbons, G. W. and Werner, M. C.",
    title = "{Applications of the Gauss-Bonnet theorem to gravitational lensing}",
    eprint = "0807.0854",
    archivePrefix = "arXiv",
    primaryClass = "gr-qc",
    doi = "10.1088/0264-9381/25/23/235009",
    journal = "Class. Quant. Grav.",
    volume = "25",
    pages = "235009",
    year = "2008"
}

@article{Takizawa:2020egm,
    author = "Takizawa, Keita and Ono, Toshiaki and Asada, Hideki",
    title = "{Gravitational deflection angle of light: Definition by an observer and its application to an asymptotically nonflat spacetime}",
    eprint = "2001.03290",
    archivePrefix = "arXiv",
    primaryClass = "gr-qc",
    doi = "10.1103/PhysRevD.101.104032",
    journal = "Phys. Rev. D",
    volume = "101",
    number = "10",
    pages = "104032",
    year = "2020"
}

@article{Islam:1983rxp,
    author = "Islam, J. N.",
    title = "{The cosmological constant and classical tests of general relativity}",
    doi = "10.1016/0375-9601(83)90756-9",
    journal = "Phys. Lett. A",
    volume = "97",
    pages = "239--241",
    year = "1983"
}

@article{Rindler:2007zz,
    author = "Rindler, Wolfgang and Ishak, Mustapha",
    title = "{Contribution of the cosmological constant to the relativistic bending of light revisited}",
    eprint = "0709.2948",
    archivePrefix = "arXiv",
    primaryClass = "astro-ph",
    doi = "10.1103/PhysRevD.76.043006",
    journal = "Phys. Rev. D",
    volume = "76",
    pages = "043006",
    year = "2007"
}

@article{Gibbons:2008ru,
    author = "Gibbons, G. W. and Warnick, C. M. and Werner, M. C.",
    title = "{Light-bending in Schwarzschild-de-Sitter: Projective geometry of the optical metric}",
    eprint = "0808.3074",
    archivePrefix = "arXiv",
    primaryClass = "gr-qc",
    reportNumber = "DAMTP-2008-74",
    doi = "10.1088/0264-9381/25/24/245009",
    journal = "Class. Quant. Grav.",
    volume = "25",
    pages = "245009",
    year = "2008"
}

@article{Arakida:2017hrm,
    author = "Arakida, Hideyoshi",
    title = "{Light deflection and Gauss{\textendash}Bonnet theorem: definition of total deflection angle and its applications}",
    eprint = "1708.04011",
    archivePrefix = "arXiv",
    primaryClass = "gr-qc",
    doi = "10.1007/s10714-018-2368-2",
    journal = "Gen. Rel. Grav.",
    volume = "50",
    number = "5",
    pages = "48",
    year = "2018"
}

@article{Hu:2021yzn,
    author = "Hu, Lingyi and Heavens, Alan and Bacon, David",
    title = "{Light bending by the cosmological constant}",
    eprint = "2109.09785",
    archivePrefix = "arXiv",
    primaryClass = "astro-ph.CO",
    doi = "10.1088/1475-7516/2022/02/009",
    journal = "JCAP",
    volume = "02",
    number = "02",
    pages = "009",
    year = "2022"
}

@article{Takizawa:2021jxa,
    author = "Takizawa, Keita and Asada, Hideki",
    title = "{Gravitational lens on de Sitter background}",
    eprint = "2112.00311",
    archivePrefix = "arXiv",
    primaryClass = "gr-qc",
    doi = "10.1103/PhysRevD.105.084022",
    journal = "Phys. Rev. D",
    volume = "105",
    number = "8",
    pages = "084022",
    year = "2022"
}

@article{Janis:1968zz,
    author = "Janis, Allen I. and Newman, Ezra T. and Winicour, Jeffrey",
    title = "{Reality of the Schwarzschild Singularity}",
    doi = "10.1103/PhysRevLett.20.878",
    journal = "Phys. Rev. Lett.",
    volume = "20",
    pages = "878--880",
    year = "1968"
}

@article{Fisher:1948yn,
    author = "Fisher, I. Z.",
    title = "{Scalar mesostatic field with regard for gravitational effects}",
    eprint = "gr-qc/9911008",
    archivePrefix = "arXiv",
    journal = "Zh. Eksp. Teor. Fiz.",
    volume = "18",
    pages = "636--640",
    year = "1948"
}

@article{Wyman:1981bd,
    author = "Wyman, M.",
    title = "{Static Spherically Symmetric Scalar Fields in General Relativity}",
    doi = "10.1103/PhysRevD.24.839",
    journal = "Phys. Rev. D",
    volume = "24",
    pages = "839--841",
    year = "1981"
}

@article{Virbhadra:2002ju,
    author = "Virbhadra, K. S. and Ellis, G. F. R.",
    title = "{Gravitational lensing by naked singularities}",
    doi = "10.1103/PhysRevD.65.103004",
    journal = "Phys. Rev. D",
    volume = "65",
    pages = "103004",
    year = "2002"
}

@article{Chen:2023uuy,
    author = "Chen, Deyou and Chen, Yiqian and Wang, Peng and Wu, Tianshu and Wu, Houwen",
    title = "{Gravitational lensing by transparent Janis{\textendash}Newman{\textendash}Winicour naked singularities}",
    eprint = "2309.00905",
    archivePrefix = "arXiv",
    primaryClass = "gr-qc",
    doi = "10.1140/epjc/s10052-024-12950-z",
    journal = "Eur. Phys. J. C",
    volume = "84",
    number = "6",
    pages = "584",
    year = "2024"
}

@misc{Zhang:2025hzh,
    author = "Zhang, Zeqian",
    title = "{Observational Signatures of Janis-Newman-Winicour Strongly Naked Singularity}",
    eprint = "2505.13214",
    archivePrefix = "arXiv",
    primaryClass = "gr-qc",
    month = "5",
    year = "2025"
}

@article{Perlick:2004tq,
    author = "Perlick, V.",
    title = "{Gravitational lensing from a spacetime perspective}",
    doi = "10.12942/lrr-2004-9",
    journal = "Living Rev. Rel.",
    volume = "7",
    pages = "9",
    year = "2004"
}

@book{Schneider:1992bmb,
    author = {Schneider, Peter and Ehlers, J{\"u}rgen and Falco, Emilio E.},
    title = "{Gravitational Lenses}",
    doi = "10.1007/978-3-662-03758-4",
    isbn = "978-3-540-66506-9, 978-3-662-03758-4",
    publisher = "Springer",
    series = "Astronomy and Astrophysics Library",
    year = "1992"
}

@article{Bozza:2002zj,
    author = "Bozza, V.",
    title = "{Gravitational lensing in the strong field limit}",
    eprint = "gr-qc/0208075",
    archivePrefix = "arXiv",
    doi = "10.1103/PhysRevD.66.103001",
    journal = "Phys. Rev. D",
    volume = "66",
    pages = "103001",
    year = "2002"
}

@article{Virbhadra:1998dy,
    author = "Virbhadra, K. S. and Narasimha, D. and Chitre, S. M.",
    title = "{Role of the scalar field in gravitational lensing}",
    eprint = "astro-ph/9801174",
    archivePrefix = "arXiv",
    journal = "Astron. Astrophys.",
    volume = "337",
    pages = "1--8",
    year = "1998"
}

@article{Sau:2020xau,
    author = "Sau, Subhadip and Banerjee, Indrani and SenGupta, Soumitra",
    title = "{Imprints of the Janis-Newman-Winicour spacetime on observations related to shadow and accretion}",
    eprint = "2004.02840",
    archivePrefix = "arXiv",
    primaryClass = "gr-qc",
    doi = "10.1103/PhysRevD.102.064027",
    journal = "Phys. Rev. D",
    volume = "102",
    number = "6",
    pages = "064027",
    year = "2020"
}

@article{Gyulchev:2019tvk,
    author = "Gyulchev, Galin and Nedkova, Petya and Vetsov, Tsvetan and Yazadjiev, Stoytcho",
    title = "{Image of the Janis-Newman-Winicour naked singularity with a thin accretion disk}",
    eprint = "1905.05273",
    archivePrefix = "arXiv",
    primaryClass = "gr-qc",
    doi = "10.1103/PhysRevD.100.024055",
    journal = "Phys. Rev. D",
    volume = "100",
    number = "2",
    pages = "024055",
    year = "2019"
}

@article{Claudel:2000yi,
    author = "Claudel, Clarissa-Marie and Virbhadra, K. S. and Ellis, G. F. R.",
    title = "{The Geometry of photon surfaces}",
    eprint = "gr-qc/0005050",
    archivePrefix = "arXiv",
    doi = "10.1063/1.1308507",
    journal = "J. Math. Phys.",
    volume = "42",
    pages = "818--838",
    year = "2001"
}

@misc{Huang:2025mek, author = "Huang, Yang", title = "{Applying the Gibbons-Werner method to bound orbits of massive particles in stationary spacetimes}", eprint = "2512.02806", archivePrefix = "arXiv", primaryClass = "gr-qc", month = "12", year = "2025" }

@article{Ghosh:2025jpb,
    author = "Ghosh, Aniruddha and Debnath, Ujjal",
    title = "{Gravitational lensing by deformed Horava-Lifshitz black hole in rainbow gravity}",
    doi = "10.1016/j.physletb.2025.139686",
    journal = "Phys. Lett. B",
    volume = "868",
    pages = "139686",
    year = "2025"
}

@article{Qiao:2025ojr, author = "Qiao, Chen-Kai and Su, Ping and Huang, Yang", title = "{A general discussion on photon spheres in different categories of spacetimes}", reportNumber = "ChinaXiv:202412.00090", doi = "10.1140/epjc/s10052-025-14435-z", journal = "Eur. Phys. J. C", volume = "85", number = "6", pages = "709", year = "2025" }

@article{Huang:2025vqm,
    author = "Huang, Yang and Fu, Xiangyun and Lu, Zhenyan and Qin, Xin",
    title = "{The influence of cosmological constant on light deflection in rotating spacetimes via the generalized Gibbons{\textendash}Werner method}",
    eprint = "2505.06999",
    archivePrefix = "arXiv",
    primaryClass = "gr-qc",
    doi = "10.1016/j.dark.2025.102070",
    journal = "Phys. Dark Univ.",
    volume = "50",
    pages = "102070",
    year = "2025"
}

@article{Lu:2025mcm,
    author = "Lu, Yi and Pan, Xiao-Yin and Lai, Meng-Yun and Wang, Qing-hai",
    title = "{Finite-distance gravitational lensing of a global monopole in a Schwarzschild{\textendash}de Sitter spacetime}",
    eprint = "2504.00777",
    archivePrefix = "arXiv",
    primaryClass = "gr-qc",
    doi = "10.1103/spt9-mpg2",
    journal = "Phys. Rev. D",
    volume = "112",
    number = "6",
    pages = "064051",
    year = "2025"
}

@article{Mushtaq:2025ksk, author = "Mushtaq, Farzan and Jawad, Abdul and Tiecheng, Xia and Alam, Mohammad Mahtab and Shaymatov, Sanjar", title = "{Probing deflection angle inspired by weak field of specific black holes in non plasma and plasma field}", doi = "10.1016/j.dark.2025.101872", journal = "Phys. Dark Univ.", volume = "48", pages = "101872", year = "2025" }

@article{Wang:2025cyd,
    author = "Wang, Qianchuan and Jia, Junji",
    title = "{Weak deflection angle of charged signal in magnetic fields}",
    eprint = "2501.03554",
    archivePrefix = "arXiv",
    primaryClass = "gr-qc",
    doi = "10.1088/1475-7516/2025/05/074",
    journal = "JCAP",
    volume = "05",
    pages = "074",
    year = "2025"
}

@article{Pantig:2024asu,
    author = {Pantig, Reggie C. and Lambiase, Gaetano and {\"O}vg{\"u}n, Ali and Lobos, Nikko John Leo S.},
    title = "{Spacetime-curvature induced uncertainty principle: Linking the large-structure global effects to the local black hole physics}",
    eprint = "2412.00303",
    archivePrefix = "arXiv",
    primaryClass = "gr-qc",
    doi = "10.1016/j.dark.2025.101817",
    journal = "Phys. Dark Univ.",
    volume = "47",
    pages = "101817",
    year = "2025"
}

@article{Pantig:2024kqy,
    author = "Pantig, Reggie C.",
    title = "{On the analytic generalization of particle deflection in the weak field regime and shadow size in light of EHT constraints for Schwarzschild-like black hole solutions}",
    eprint = "2409.00476",
    archivePrefix = "arXiv",
    primaryClass = "gr-qc",
    doi = "10.1140/epjc/s10052-025-13766-1",
    journal = "Eur. Phys. J. C",
    volume = "85",
    number = "1",
    pages = "52",
    year = "2025"
}

@article{Gao:2024ejs,
    author = "Gao, Xiao-Jun",
    title = "{Gravitational lensing and shadow by a Schwarzschild-like black hole in metric-affine bumblebee gravity}",
    eprint = "2409.12531",
    archivePrefix = "arXiv",
    primaryClass = "gr-qc",
    doi = "10.1140/epjc/s10052-024-13338-9",
    journal = "Eur. Phys. J. C",
    volume = "84",
    number = "9",
    pages = "973",
    year = "2024"
}

@article{Liu:2024wal,
    author = "Liu, Hao and Lai, Meng-Yun and Pan, Xiao-Yin and Huang, Hyat and Zou, De-Cheng",
    title = "{Gravitational lensing effect of black holes in effective quantum gravity}",
    eprint = "2408.11603",
    archivePrefix = "arXiv",
    primaryClass = "gr-qc",
    doi = "10.1103/PhysRevD.110.104039",
    journal = "Phys. Rev. D",
    volume = "110",
    number = "10",
    pages = "104039",
    year = "2024"
}

@article{Zhang:2024uex,
    author = "Zhang, Zhen and Zhang, Rui",
    title = "{On the global Gaussian bending measure and its applications in stationary spacetimes}",
    eprint = "2408.02195",
    archivePrefix = "arXiv",
    primaryClass = "gr-qc",
    doi = "10.1088/1361-6382/add788",
    journal = "Class. Quant. Grav.",
    volume = "42",
    number = "11",
    pages = "115006",
    year = "2025"
}

@article{Ovgun:2025ctx,
    author = {{\"O}vg{\"u}n, Ali},
    title = "{Weak gravitational lensing in Ricci-coupled Kalb{\textendash}Ramond bumblebee gravity: Global monopole and axion-plasmon medium effects}",
    eprint = "2504.07130",
    archivePrefix = "arXiv",
    primaryClass = "gr-qc",
    doi = "10.1016/j.dark.2025.101905",
    journal = "Phys. Dark Univ.",
    volume = "48",
    pages = "101905",
    year = "2025"
}

@article{Soares:2025hpy,
    author = "Soares, A. R. and Pereira, C. F. S. and Vit{\'o}ria, R. L. L. and Silva, Marcos V. de S. and Belich, H.",
    title = "{Light deflection and gravitational lensing effects inspired by loop quantum gravity}",
    eprint = "2503.06373",
    archivePrefix = "arXiv",
    primaryClass = "gr-qc",
    doi = "10.1088/1475-7516/2025/06/034",
    journal = "JCAP",
    volume = "06",
    pages = "034",
    year = "2025"
}

@article{Rincon:2024won,
    author = {Rinc{\'o}n, {\'A}ngel and {\"O}vg{\"u}n, Ali and Pantig, Reggie C.},
    title = "{An effective model for the quantum Schwarzschild black hole: Weak deflection angle, quasinormal modes and bounding of greybody factor}",
    eprint = "2409.10930",
    archivePrefix = "arXiv",
    primaryClass = "gr-qc",
    doi = "10.1016/j.dark.2024.101623",
    journal = "Phys. Dark Univ.",
    volume = "46",
    pages = "101623",
    year = "2024"
}

@article{Li:2024ujw,
    author = "Li, Zonghai",
    title = "{Gravitational lensing using Werner{\textquoteright}s method in Cartesian-like coordinates}",
    eprint = "2404.19658",
    archivePrefix = "arXiv",
    primaryClass = "gr-qc",
    doi = "10.1103/PhysRevD.111.084017",
    journal = "Phys. Rev. D",
    volume = "111",
    number = "8",
    pages = "084017",
    year = "2025"
}

@article{Liu:2023xtb,
    author = "Liu, Yi-Gao and Qiao, Chen-Kai and Tao, Jun",
    title = "{Gravitational lensing of spherically symmetric black holes in dark matter halos}",
    eprint = "2312.15760",
    archivePrefix = "arXiv",
    primaryClass = "gr-qc",
    doi = "10.1088/1475-7516/2024/10/075",
    journal = "JCAP",
    volume = "10",
    pages = "075",
    year = "2024"
}

@article{Gao:2023ltr,
    author = "Gao, Xiao-Jun and Yan, Xiao-kun and Yin, Yihao and Hu, Ya-Peng",
    title = "{Gravitational lensing by a charged spherically symmetric black hole immersed in thin dark matter}",
    eprint = "2303.00190",
    archivePrefix = "arXiv",
    primaryClass = "gr-qc",
    doi = "10.1140/epjc/s10052-023-11414-0",
    journal = "Eur. Phys. J. C",
    volume = "83",
    number = "4",
    pages = "281",
    year = "2023"
}

\end{document}